\documentclass{preprint}
\usepackage{times}
\usepackage{ScriptFonts}
\usepackage{myarray}
\usepackage{Martin}
\usepackage{Symbols}
\usepackage[righttag]{MHequ}
\usepackage{MHenvs}
\usepackage{sub_JP}
\def\IL{\I_{\!\rL}}
\def\CPR{\CP_{\!\rho}}
\def\FR{F_{\!\rho}}
\def\PhiR{\Phi_{\rho}}
\def\u#1{u^{(#1)}}
\def\L{{\rm L}}
\def\tT{\widetilde T}
\def\YY#1#2{{\cal G}^{#1}_{#2}}
\def\XX#1#2{\CH^{#1}_{#2}}
\def\PT#1{\Phi^{#1}}
\def\SD{\cscr D}
\def\SL{\cscr L}
\def\SA{\cscr A}
\def\SH{{\cal H}}
\def\CHH{\CH}
\def\WSH{\widetilde{\SH}}
\def\SB{\cscr B}
\def\hF{\widehat F}
\def\tF{\widetilde F}
\def\hA{{\widetilde A}}
\def\hQ{{\widetilde Q}}

\def\rL{{\rm L}}
\def\rH{{\rm H}}
\def\tr{\mop{Tr}}
\def\poly{P}

\def\CWP{\CW_{\hbox{\scriptsize \rm
per}}^{1}([-\pi,\pi])}
\def\ic{\xi} 
\def\vzero{h_0}
\def\uzero{h}
\usefont{OT1}{cmtt}{m}{n}
\fontsize{10}{10}
\selectfont
\expandafter\fontdimen4\csname OT1/cmtt/m/n/10\endcsname=2pt
\expandafter\fontdimen3\csname OT1/cmtt/m/n/10\endcsname=2pt
\usefont{OT1}{cmtt}{m}{n}
\fontsize{12}{12}
\selectfont
\expandafter\fontdimen4\csname OT1/cmtt/m/n/12\endcsname=2pt
\expandafter\fontdimen3\csname OT1/cmtt/m/n/12\endcsname=2pt
\def\ttt#1{{\hfill\break\null\kern
-2truecm
{\tt ********** #1}
}} 
\def\Exp{{\symb E}}
\def\Prob{{\symb P}}
\def\norm {|\!|\!|}
\def\red#1{
#1
}

\def\I{\cscr I}
\def\ie{{\it i.e.}}
\def\eg{{\it e.g.}}
\def\kmax{L}
\def\dimI{{W}}
\def\transpose{*}
\def\HALF{{\textstyle{1\over 2}}}
\def\myitem#1{{\rm (#1)}}
\def\kzero{k_*}

\def\cp{\circ\Phi^t}
\def\omit#1{}
\let\epsilon=\varepsilon

\def\SY{{\cscr Y}}

\def\SXb{\tilde {\cscr X \kern3pt}\kern-3pt}
\def\HS{{\hbox{\scriptsize{\rm HS}}}}
\def\sumiL{\sum_{i=0}^{L-1}}
\def\threebars{|\!|\!|}
\def\thrb{\big|\!\big|\!\big|}
\def\thrB{\Big|\!\Big|\!\Big|}
\def\thrbb{\bigg|\!\bigg|\!\bigg|}
\def\thrBB{\Bigg|\!\Bigg|\!\Bigg|}
\renewcommand\n[3][ ]{%
\ifthenelse{\equal{#1}{ }}{|{#3}|_{#2}}{}%
\ifthenelse{\equal{#1}{b}}{\bigl|{#3}\bigr|_{#2}}{}%
\ifthenelse{\equal{#1}{B}}{\Bigl|{#3}\Bigr|_{#2}}{}%
\ifthenelse{\equal{#1}{bb}}{\biggl|{#3}\biggr|_{#2}}{}%
\ifthenelse{\equal{#1}{BB}}{\Biggl|{#3}\Biggr|_{#2}}{}}%
\newcommand\nn[3][ ]{%
\ifthenelse{\equal{#1}{ }}{\|{#3}\|_{#2}}{}%
\ifthenelse{\equal{#1}{b}}{\bigl\|{#3}\bigr\|_{#2}}{}%
\ifthenelse{\equal{#1}{B}}{\Bigl\|{#3}\Bigr\|_{#2}}{}%
\ifthenelse{\equal{#1}{bb}}{\biggl\|{#3}\biggr\|_{#2}}{}%
\ifthenelse{\equal{#1}{BB}}{\Biggl\|{#3}\Biggr\|_{#2}}{}}%
\newcommand\nnn[3][ ]{%
\ifthenelse{\equal{#1}{ }}{\threebars{#3}\threebars_{#2}}{}%
\ifthenelse{\equal{#1}{b}}{\mathopen\thrb{#3}\mathclose\thrb_{#2}}{}%
\ifthenelse{\equal{#1}{B}}{\mathopen\thrB{#3}\mathclose\thrB_{#2}}{}%
\ifthenelse{\equal{#1}{bb}}{\mathopen\thrbb{#3}\mathclose\thrbb_{#2}}{}%
\ifthenelse{\equal{#1}{BB}}{\mathopen\thrBB{#3}\mathclose\thrBB_{#2}}{}}%
\def\defcommand#1:#2:{\global\let\myref\myundefined%
\ifthenelse{\equal{#1}{e}}{\global\let\myref\eref}{}%
\ifthenelse{\equal{#1}{prop}}{\global\let\myref\prop}{}%
\ifthenelse{\equal{#1}{lem}}{\global\let\myref\lem}{}%
\ifthenelse{\equal{#1}{rem}}{\global\let\myref\rem}{}%
\ifthenelse{\equal{#1}{theo}}{\global\let\myref\theo}{}%
\ifthenelse{\equal{#1}{sec}}{\global\let\myref\sect}{}%
\ifthenelse{\equal{#1}{def}}{\global\let\myref\mydefinition}{}%
}
\def\mydefinition#1{Definition~\ref{#1}}
\def\myundefined#1{\red{whatsthis #1?}}
\def\sref#1{\defcommand#1:%
\myref{#1}}
\def\inflikesup{\mathop{\rm inf{\vphantom{p}}}}
\let\Ann=\nn
\let\Bnn=\nn
\begin{document}

\title{Uniqueness of the Invariant Measure for a\\
Stochastic PDE Driven by Degenerate Noise}
\author{J.-P.~Eckmann\inst{1}\fnmsep\inst{2}, M.~Hairer\inst{1}}
\institute{D\'epartement de Physique Th\'eorique, Universit\'e de Gen\`eve \and
Section de Math\'ematiques, Universit\'e de Gen\`eve
 \\ \email{Jean-Pierre.Eckmann@physics.unige.ch}\\
\email{Martin.Hairer@physics.unige.ch}}
\titleindent=0.65cm

\maketitle
\thispagestyle{empty}
\mynumbering
\pagestyle{MHheadings}

\begin{abstract}
We consider the stochastic Ginzburg-Landau equation in a bounded
domain.
We assume the stochastic forcing acts only on high spatial
frequencies. The low-lying frequencies are then only connected to this
forcing through the non-linear (cubic) term of the Ginzburg-Landau equation.
Under these assumptions, we show that the stochastic PDE has a {\em
unique} invariant measure.
The techniques of proof combine a controllability argument for the
low-lying frequencies with an infinite dimensional version of the
Malliavin calculus to show positivity and regularity of the invariant
measure. This then implies the uniqueness of that measure.
\end{abstract}

\tableofcontents

\section{Introduction}
In this paper, we study a stochastic variant of the Ginzburg-Landau
equation on a finite domain with periodic boundary conditions.
The deterministic equation is
\begin{equ}[e:GL0]
\dot u = \Delta u + u - u^3\;,\quad u(0) = \u0 \in \CH\;,
\end{equ}
where $\CH$ is the real Hilbert space $\CWP$, \ie, the closure of the space of smooth periodic
functions $u:[-\pi,\pi] \to \R$ equipped with the norm
\begin{equ}
\|u\|^2 = \int_{-\pi}^{\pi} \bigl(|u(x)|^2 + |u'(x)|^2\bigr)\,dx\;.
\end{equ}
(The restriction to the interval
$[-\pi,\pi]$ is irrelevant since other lengths of intervals can be obtained
by scaling space, time and amplitude $u$ in \eref{e:GL0}.)
While we work exclusively with the real Ginzburg-Landau equation
\eref{e:GL0}
our methods generalize immediately to the complex Ginzburg-Landau
equation
\begin{equ}[e:CGL0]
\dot u = (1+ia)\Delta u + u - (1+ib ) |u|^2 u\;,\quad a,b\in\R\;,
\end{equ}
which has a more interesting dynamics than \eref{e:GL0}. But the
notational details are slightly more involved because of the complex
values of $u$ and so we stick with \eref{e:GL0}.   

While a lot is known about existence and regularity of solutions of
\eref{e:GL0} or \eref{e:CGL0},  
only very little information has been obtained about the
attractor of such systems, and in particular, nothing seems
to be known about invariant measures on the attractor. 

On the other hand, when
\eref{e:GL0} is replaced by a stochastic differential equation, more
can be said about the invariant measure, see \cite{ZDP} and references therein.
Since the problem \eref{e:GL0} involves only functions with periodic boundary
conditions, it can be rewritten in terms of the Fourier series for
$u$:
$$
u(x,t)\,=\,\sum_{k\in\Z} e^{i k x} u_k(t)\;,\quad
u_k\,=\,{1\over 2\pi}\int_{-\pi}^{\pi}  e^{-ikx} u(x)\,dx\;.
$$
We call $k$ the momenta, $u_k$ the modes,
and, since $u(x,t)$ is real we must always
have $u_k(t)=\bar u_{-k}(t)$, where $\bar z$ is the complex conjugate
of $z$. With these notations
\eref{e:GL0}
takes the form
\begin{equ}
\dot u_k \,=\, (1-k^2) u_k -\sum_{k_1+k_2+k_3=k} u_{k_1} u_{k_2} u_{k_3}\;,
\end{equ}
for all $k\in\Z$ and the initial condition satisfies 
$\{(1 +|k|)u_k(0)\}\in\ell^2$. In the sequel, we will use the symbol $\CH$
indifferently for the space
$\CWP$ and for its counterpart in
Fourier
space.
In the earlier literature on uniqueness of the invariant
measure for stochastic differential equations, see
\eg, \cite{ZDP},
the authors are mostly interested in systems
where each of the $u_k$ is forced by some external noise term. The
main aim of our work is to study forcing by noise
which {\em acts  only on the high-frequency part} of $u$, namely on the $u_k$
with $|k|\ge \kzero$ for some finite $\kzero\in\N$. The low-frequency
amplitudes $u_k$ with $|k|<\kzero$ are then only {\em indirectly} forced
through the noise, namely through the nonlinear coupling of the
modes. In this respect, our approach is reminiscent of the work done on
thermally driven chains in \cite{EPR,EPR2,EH}, where the chains were only
stochastically driven at the ends.

In the context of our problem, the {\em existence} of
an
invariant measure is a classical result for the noise we consider
\cite{ZDP}, and the main novelty of our paper is a proof of {\em
uniqueness} 
of that measure.
To prove uniqueness we begin by
proving controllability of the equations, \ie, to
show that the high-frequency noise together with
non-linear coupling effectively drives the low-frequency modes. Using
this, we then
use Malliavin calculus in infinite dimensions,
to show regularity of the transition probabilities. This then implies
uniqueness of the invariant measure.

We will study the system of equations
\begin{equ}[e:SGL]
du_k = -k^2 u_k \, dt + \bigl(u_k - (u^3)_k\bigr)\,dt + {q_k\over
\sqrt{4\pi (1+k^2)}}\,dw_k(t)\;,
\end{equ}
with $u \in \CH$.
The above equations hold for $k \in \Z$, and it is always
understood that
\begin{equ}[e:cubic]
(u^3)_k=\sum_{k_1+k_2+k_3=k\atop k_1,k_2,k_3\in\Z} u_{k_1}
u_{k_2} u_{k_3}\;,
\end{equ}
with $u_{-k}=\bar u_k$.
{\em To avoid inessential notational problems we will work with even
periodic functions, so that $u_k=u_{-k}\in\R$}.
We will work with the basis 
\begin{equ}[e:basis]
e_k(x)\,=\,{1\over \sqrt{\pi (1+k^2)}} \cos(kx)\;.
\end{equ}
Note that this basis is orthonormal w.r.t.~the scalar product in
$\CH$, but the $u_k$ are actually given by
$u_k=(4\pi(1+k^2))^{-1/2}\scal{u,e_k}$. (We choose this to make the
cubic term \eref{e:cubic} look simple.)

The noise is supposed to act only on the high frequencies, but there
we need it to be strong enough in the following way. Let $a_k=k^2+1$.
Then we require
that there exist
constants $c_1,c_2 > 0$ such that for $k \ge \kzero$,
\begin{equ}[e:coeff]
c_1 a_k^{-\alpha} \le q_k \le c_2 a_k^{-\beta}\;,\qquad \alpha \ge 2\;,\quad
\alpha - 1/8 < \beta \le \alpha\;.
\end{equ}
These conditions imply
\begin{equs}
\sum_{k=0}^\infty (1+k^{4\alpha -3/2} )q_k^2 <\infty \;,\\
\sup_{k\ge\kzero}k^{-2\alpha } q_k^{-1}<\infty \;.
\end{equs}

We formulate the problem in a more general setting:
Let $F(u)$ be a polynomial of odd degree with negative leading
coefficient.
Let $A$ be the operator of multiplication by $1+k^2$ and let $Q$ be
the operator of multiplication by $q_k$. Then \eref{e:SGL} is of the
form
\begin{equ}[e:SGLL]
d\Phi^t\,=\, -A\Phi^t \,dt + F(\Phi^t)\,dt + Q \,dW(t)\;,
\end{equ}
where $dW(t)=\sum_{k=0}^\infty e_k dw_k(t)$ is the cylindrical Wiener
process on $\CH$ with the $w_k$ mutually independent real Brownian motions.\footnote{It 
is convenient to have, in the case of \eref{e:SGL}, $A=1-\Delta$ and
$F(u)=2u-u^3$ rather than $A=-1-\Delta$ and $F(u)=-u^3$.}
We define $\Phi^t(\xi)$ as the solution of \eref{e:SGLL} with initial
condition $\Phi^0(\xi)=\xi$.
Clearly, the conditions on $Q$ can be formulated as
\minilab{e:allerletztes}
\begin{equs}[0]
\nn{\HS} {A^{\alpha -3/8} Q}<\infty \;,\label{e:zzz1}\\
q_k^{-1}k^{-2\alpha } \text { is bounded for $k\ge\kzero $}\;,\label{e:zzz2}
\end{equs}
where $\nn{\HS}{\,\cdot\,}$ is the Hilbert-Schmidt norm on $\CH$. Note
that for each $k$, \eref{e:SGL} is obtained by multiplying
\eref{e:SGLL} by $(4\pi(1+k^2))^{-1/2}\scal{\cdot,e_k}$.


\begin{likerem}[Important Remark.]
The crucial aspect of our conditions
is the possibility of choosing $q_k=0$ for all $k<\kzero $, \ie, the noise
drives only the high frequencies. But we also allow any of the $q_k$
with $k<k_*$ to be different from 0, which corresponds to long
wavelength forcing. Furthermore, as we are allowing
$\alpha $ to be arbitrarily large, this means that the forcing at high
frequencies has an amplitude which can decay like any power. The point
of this paper is to show that these conditions are sufficient to
ensure the existence of a unique invariant measure for \eref{e:SGLL}.
\end{likerem}

\begin{theorem}
\label{theo:super}The process \eref{e:SGLL} has a unique invariant Borel
measure on $\CH$. 
\end{theorem} 

There are two main steps in the proof of \theo{theo:super}. First, 
the nature of the nonlinearity $F$ implies that the modes with
$k\ge \kzero $ couple in such a way to those with $k<\kzero $ as to allow {\em
controllability}. Intuitively, this means that any point in phase
space can be reached to arbitrary precision in any given time, by a
suitable choice of the high-frequency controls.

Second, we show that a version of the Malliavin calculus can be
implemented in our infinite-dimensional context. This will be the hard
part of our study, and the main result of that part is a proof that
the
strong Feller property holds. This means that
for any measurable function $\phi\in\CB_b(\CH)$, the function
\begin{equ}[e:PTdef]
\bigl(\CP^t \phi\bigr)(\xi)\,\equiv\,\Exp\Bigl(
\bigl(\phi\circ \Phi^t\bigr)(\xi)
\Bigr)
\end{equ}
is {\em continuous}.\footnote{Throughout the paper, $\Exp$ 
denotes expectation and $\Prob$ denotes probability for the random
variables.}
We show this by proving that a
cutoff version of \eref{e:SGLL} (modifying the dynamics at large
amplitudes by a parameter $\rho$) makes $\CPR^t  \phi$ a
{\em differentiable} map.

The interest in such highly degenerate stochastic PDE's is related to
questions in hydrodynamics
where one would 
ask how ``energy'' is transferred from high to low
frequency modes, and vice versa when only some of the modes are
driven.
This could then shed some light on
the entropy-enstrophy problem in the (driven) Navier-Stokes equation.

To end this introduction, we will try to compare the results of our paper
to current work of others. 
These groups consider the $2$-D Navier Stokes equation without
deterministic external forces, also in bounded domains. 
In these equations, any initial condition eventually converges to
zero, as long as there is no stochastic forcing. 
First there is earlier work by Flandoli-Maslovski \cite{FM} dealing with noise 
whose amplitude is bounded below by $\vert k\vert^{-c}$. 
In the work of \cite{BKL}, 
the stochastic forcing acts on modes with low
$k$, and they get uniqueness of the invariant measure and analyticity, with
probability 1. 
In the work of Kuksin and Shirikyan \cite{KS00} the bounded noise
is quite general acts on all Fourier modes, and acts at definite times
with "noise-less" intervals in-between. 
Again, the measure is unique.
It is supported by $\CC^\infty $ functions, is mixing and has a Gibbs property.
In the work of \cite{EMS}, uniqueness is shown for NS by forcing only 3 modes.

The main difference between those results and the present paper is our
control of a situation which is already unstable at the deterministic
level. 
Thus, in this sense, it comes closer to a description of a
deterministically turbulent fluid (\eg, obtained by an external
force). 
On the other hand, in our work, we need to actually force all high spatial
frequencies. 
Perhaps, this could be eliminated by a combination with ideas from the
papers above. 

\section{Some Preliminaries on the Dynamics}

Here, we summarize some facts about deterministic and stochastic GL
equations from the literature which we need to get started.

We will consider the dynamics on the following space:
\begin{definition}
We define
$\CH$ as the subspace of even functions in $\CWP$.
The norm on $\CH$ will be denoted by $\|\,\cdot\,\|$\,, and the scalar
product
by $\scal{\cdot,\cdot}$\,.
\end{definition}

We consider first the deterministic equation 
\begin{equ}[e:GL]
\dot u = \Delta u + u - u^3\;,\quad u(0) = \u0 \in \CH\;,
\end{equ}
Due to its dissipative character
the solutions are, for positive times, analytic in a strip
around
the real axis. More precisely, denote by $\|\cdot\|_{\SA _\eta}$ the norm
\begin{equ}
\|f\|_{\SA_\eta} = \sup_{|\Im z|\le \eta} |f(z)|\;,
\end{equ}
and by $\cscr A_\eta$ the corresponding Banach space of analytic functions.
Then the following result holds.
\begin{lemma}
\label{lem:anal}
For every initial value $\u0 \in \CH$, there exist a time $T$ and a constant
$C$ such that for $0<t\le T$, the solution $u(t,\u0)$ of \eref{e:GL} belongs
to $\SA _{_{\!\!\sqrt t}}$ and satisfies
$\|u(t,\u0)\|_{\SA_{\sqrt{t}}} \le C$. 
\end{lemma}

\begin{proof}
The statement is proven in \cite{PC} for the case of the infinite
line. 
Since the periodic functions form an invariant subspace under the
evolution, the result applies to our case.
\end{proof}
We next collect some useful results for the stochastic equation \eref{e:SGLL}:
\begin{proposition}
\label{prop:allerlei}For every $t>0$ and every $p\ge1$ 
the solution of \eref{e:SGLL} with initial condition $\Phi^0(\xi)=\xi\in\CH$
exists in $\CH$ up to time $t$.
It defines by \eref{e:PTdef} a Markovian transition semigroup on $\CH$.
One has the bound
\begin{equ}
\Exp \Bigl(\sup_{s \in [0,t]} \|\Phi^s(\xi)\|^p\Bigr) \le C_{t,p}(1+\|\xi\|)^{p}\;.
\end{equ}
Furthermore, the process \eref{e:SGLL} has an invariant measure.
\end{proposition}
These results are well-known and in \sect{sec:theend} we sketch where
to find them in the literature.

\section{Controllability}\label{sec:Controllability}

In this section we show the ``approximate controllability'' of
\eref{e:SGL}.
The control problem under consideration is
\begin{equ}[e:control]
\dot u = \Delta u + u - u^3 + Q\,f(t)\;,\qquad u(0) = \u{\rm i} \in \CH\;,
\end{equ}
where $f$ is the control.
Using Fourier series' and the hypotheses on $Q$, we see that by
choosing $f_k\equiv0$ for $|k|<k_*$,
\eref{e:control} 
can be brought to the form
\begin{equ}[e:Fourier]
\dot u_k = \left\{\begin{array}{Ll}
-k^2 u_k + u_k - \sum_{\ell+m+n = k} u_\ell u_m u_n+
{q_k\over \sqrt{4\pi(1+k^2)}} f_k(t)\;,&\quad |k| \ge \kzero \;, \\
-k^2 u_k + u_k - \sum_{\ell+m+n = k} u_\ell u_m u_n \;, &\quad |k| < \kzero \;,
\end{array}\right.
\end{equ}
with $\{u_k\} \in \CH$ and $t \mapsto \{f_k(t)\} \in
\L^\infty([0,\tau],\CH)$.
We will refer in the sequel to $\{u_k\}_{|k| < \kzero }$ as the {\em
low-frequency
modes} and to
$\{u_k\}_{|k| \ge \kzero }$ as the {\em high-frequency modes}. We also introduce the
projectors $\Pi_{\rL}$ and $\Pi_{\rH}$ which project onto the low (resp. high)
frequency modes. Let  $\CH_{\rL}$ and $\CH_{\rH}$ denote the ranges of
$\Pi_{\rL}$ and
$\Pi_{\rH}$ respectively. Clearly $\CH_{\rL}$ is finite dimensional, whereas
$\CH_{\rH}$ is a separable Hilbert space. 

The main result of this section is approximate controllability in the
following sense: 
\begin{theorem}
\label{theo:control}
For every time $\tau > 0$ the following is true:
For every  $\u{\rm i}, \u{\rm f}  \in \CH$
and every $\eps >0$,
there exists
a control $f \in \L^\infty([0,\tau],\CH)$ such that the solution $u(t)$ of
\eref{e:control} with $u(0) = \u{\rm i}$ satisfies $\|u(\tau) - \u{\rm f} \| \le \eps$.
\end{theorem}
\begin{proof}The construction of the control proceeds in 4 different
phases, of which the third is the actual controlling of the
low-frequency part by the high-frequency controls.
In the construction, we will encounter a time $\tau (R,\epsilon ')$
which depends on the norm $R$ of $\u{\rm f}$ and some precision $\epsilon
'$. Given this function, we split the given time $\tau $ as 
$\tau =\sum_{i=1}^4 \tau _i$, with $\tau _4\le\tau
(\nn{}{\u{\rm f}},\epsilon/2)$ and all $\tau _i>0$.
We will use the cumulated times $t_j=\sum_{i=1}^j \tau _i$.

\begin{likerem}[Step 1.] In this step we choose $f\equiv 0$, and we
define
$\u1=u(t _1)$, where $t\mapsto u(t )$ is the solution of \eref{e:control}
with initial condition $u(0)=\u{\rm i}$. Since there is no control, we
really have \eref{e:GL} and hence, by \lem{lem:anal}, we see that
$\u1\in\SA_\eta$ for some $\eta>0$.
\end{likerem}
\begin{likerem}[Step 2.] We will construct a smooth control 
$f:[t_1,t _2] \to \CH$ such that $\u2=u(t _2)$ satisfies:
\begin{equ}
\Pi_\rH \u2 \,=\,0\;.
\end{equ}
In other words, in this step, we drive the high-frequency part to $0$.
To construct $f$, we 
choose a $\CC^\infty$ function $\phi : [t_1,t_2] \to \R$,
interpolating between $1$ and $0$ with vanishing derivatives at the ends.
Define 
$u_{\rH}(t) = \phi(t)\Pi_{\rH} \u1$ for $t\in[t_1,t_2]$.
This will be the evolution of
the high-frequency part.
We next define
the low-frequency part
$u_{\rL}=u_{\rL}(t)$ as the solution of the 
ordinary differential equation
\begin{equ}
\dot u_{\rL} = \Delta u_{\rL} + u_{\rL} - \Pi_{\rL}\bigl((u_{\rL} +
u_{\rH})^3\bigr)\;,
\end{equ}
with $u_\rL(t_1)=\Pi_\rL\u1$. We then set 
$u(t) = u_{\rL}(t) \oplus u_{\rH}(t)$ and substitute into
\eref{e:control} which we need to solve for the control $Qf(t)$ for
$t\in[t_1,t_2]$.

Since $u_{\rL}(t) \oplus u_{\rH}(t)$ as constructed above is in
$\SA_\eta$ and since
$Qf=\dot u -\Delta u -u + u^3$, and $\Delta$ maps $\SA_\eta$ to
$\SA_{\eta/2}$ we conclude that $Qf\in\SA_{\eta/2}$.
By construction, the components $q_k$ of $Q$ decay polynomially with
$k$ and do not vanish for $k\ge \kzero $. Therefore, $Q^{-1}$ is a bounded
operator
from $\SA _{\eta/2} \cap
\CH_{\rH}$ to $\CH_\rH$.
Thus, we can solve for $f$ in this step.
\end{likerem}
\begin{likerem}[Step 3.] As mentioned before,
this step really exploits the coupling between high and low frequencies.
Here, we start from $\u2$ at time $t_2$ and we
want to reach $\Pi_\rL \u{\rm f}$ at time $t_3$. 
In fact, we will instead reach a point $\u3$ with
$\nn{}{\Pi_\rL \u3-\Pi_\rL \u{\rm f}}<\epsilon/2$.

The
idea is to
choose for every low frequency $|k| < \kzero $ a set of
three\footnote{The number $3$ is the highest power of the
nonlinearity $F$ in the GL equation.}
high frequencies
that will be used to control $u_k$. 
\begin{definition}
\label{def:i}
We assign to every $k$ with $|k|< \kzero $
a set $\I_k \subset \{k\;:\;|k| \ge \kzero \}$ of three indices. The sets
$\I_k$ are disjoint for different $k$. We also define
$\IL = \{k\;:\;|k| < \kzero \}$ and
\begin{equ}
\I = \IL  \cup \Bigl(\bigcup_{|k| < \kzero } \!\I_k\Bigr)\;.
\end{equ}
The $\I_k$ will be constructed in such a way
that they satisfy the following conditions.
\begin{claim}
\item[\myitem{A}] Let $\I_k=\{ k_1, k_2, k_3\}$. Then, $k_1+k_2+k_3=k$,
and $|k_i|\ge\kzero $ for $i=1,2,3$.
\item[\myitem{B}] $\I_{-k} = -\I_k$, where $-\I_{k}=\{ k ~:~ -k\in \I_k\}$
(we do not require this for $k=0$).
\item[\myitem{C}] Let $S$ be a collection of three indices in $\I$, $S =
\{k_1,k_2, k_3\}$ and assume $k_1+k_2+k_3 = k$ and
$|k| < \kzero $, then either $S = \I_k$, or $S\subset\IL$, or $S$ is of the form $S = \{k, k',-k'\}$.
\end{claim}
\end{definition}
The construction of the $\I_k$ is easy and there are many
possibilities. For example, assume $\kzero =1000$. Then we choose
$\I_0 = \{ 3'000'000, 2'000'000, -5'000'000\}$. Next choose
$\I_1=\{3'001'001, 2'001'000, -5'002'000\} $. Clearly, these choices
satisfy \myitem{A} above, and all combination frequencies other than $0$ or
$1$ lie outside the set $\{ |k|<1000\}$. Defining
$\I_k=\{3'000'000+1001k, 2'000'000+1000k, -5'000'000-2000k\}$, for
$0\le k<1000$, and then $\I_{-k}=-\I_k$ for $k\ne0$, we complete the
construction. 
The generalization to arbitrary $\kzero $ is left to the reader.

We are going to construct a control which, in addition to driving the
low frequency part as indicated, also
implies
$u_k(t) \equiv 0$ for $k \not\in \I$ for $t\in[t_2,t_3]$. 
By the conditions on $\I$,
the low-frequency part of
\eref{e:Fourier} is then equal to (having chosen the
controls equal to 0 for $k< \kzero $):
\begin{equ}[e:low]
\dot u_k = \Bigl(1-k^2- 3\sum_{n \in \I \setminus \IL } |u_n|^2\Bigr) u_k -
\sum_{\ell + m + n = k \atop \{\ell,m,n\} \subset \IL } u_\ell u_m u_n -
\prod_{n \in \I_k} u_n\;.
\end{equ}
(For $k=0$ there is a factor 2 in front of the last product.)
This expression is not easy to work with. 
To simplify the combinatorial problem, we choose the
controls of the 3
amplitudes $u_n$ with $n\in\I_k$ in such a way that these
$u_n$ are all equal to a fixed
real\footnote{It is here that our choice of even functions $u$
somewhat simplifies the discussion. In the general case, one would
have to argue with complex functions $z_k$, but there are no
difficulties with this.} function $z_k(t)$ which we will determine below.
With this particular choice, \eref{e:low} reduces for $|k|< \kzero $ to
\begin{equ}[e:low2]
0=-\dot u_k + \Bigl(1-k^2- 3\sum_{|n| < \kzero } |z_n|^2 \Bigr) u_k -
\bigl((\Pi_{\rL}
u)^3\bigr)_k - z_k^3\;.
\end{equ}
We now claim that for every path $\gamma \in \CC^\infty([t_2,t_3];\CH_{\rL})$
and
every $\eps>0$, {\em we can find a set of bounded functions $t\mapsto
z_k(t)$ 
such that the solution of
\eref{e:low2} shadows $\gamma$ at a distance at most $\eps$.}

To prove this statement, consider the map $F : \R^{k_*} \to \R^{k_*}$ of
the form (obtained when substituting the path $\gamma $ into \eref{e:low2})
\begin{equ}
F : \pmatrix{z_0\cr z_1\cr  \vdots\cr z_{k_*-1}\cr} \mapsto
\pmatrix{F_0(z)\cr F_1(z)\cr\vdots\cr F_{{k_*-1}}(z)} = 
\pmatrix{2z_0^3  \cr z_1^3~~  \cr\vdots\cr z_{k_*-1}^3} +
\pmatrix{\CP_0(z)\cr\CP_1(z)\cr\vdots\cr\CP_{{k_*-1}}(z)}\;,
\end{equ}
where the $\CP_m$ are polynomials of degree at most $2$. We want to
find a solution to $F=0$.
The $F_m$ form a Gr\"obner basis for the ideal of the ring of
polynomials they generate. As an immediate consequence, the
equation $F(z) = 0$ possesses exactly $3^{{k_*}}$ complex solutions, if they are
counted with multiplicities \cite{MSt}. Since the coefficients of
the $\CP_m$ are real this implies that there exists at
least one real 
solution.

Having found a (possibly discontinuous) solution for the $z_k$, we find
nearby smooth functions $\tilde z_k$ with the following properties:
\begin{claim}
\item[--]The equation \eref{e:low2} with $\tilde z_k$ replacing
$z_k$
and initial condition $u_k(t_2)=\u2_k$ leads to a solution $u$ with
$\nn{}{u(t_3)-\Pi_\rL \u{\rm f}}\,\le\,\epsilon /2$.
\item[--]One has $\tilde z_k(t_3)=0$.
\end{claim}
Having found the $\tilde z_k$ we construct the $f_k$ in such a way
that for $n\in\I_k$ one has $u_n(t)=\tilde z_k(t)$. Finally, for $k\notin \I$
we choose the controls in such a way that $u_k(t)\equiv0$ for
$t\in[t_2, t_3]$. 
We define $\u3$ as the solution obtained in this way for $t=t_3$.
\end{likerem}
\begin{likerem}[Step 4.]  Starting from $\u3$ we want to reach
$\u{\rm f}$. Note that $\u3$ is in $\SA_\eta$ (for every $\eta>0$)
since it has only a finite
number of non-vanishing modes. By construction we also have 
$\nn{}{\Pi_\rL\u3-\Pi_\rL \u{\rm f}}\,\le\,\epsilon /2$. We only need to
adapt the high frequency part without moving the low-frequency part
too much.

Since $\SA_\eta$ is dense in $\CH$, there is a $
\u4\in\SA_\eta$ with $\nn{}{
\u4-\u{\rm f}}\,\le\,\epsilon /4$. By the reasoning of Step 2
there is for every $\tau ' >0$
a control for which $\Pi_\rH u(t_3+\tau ' )=\Pi_\rH\u4$ when
starting from $u(t_3)=\u3$. Given $\epsilon $ there is a $\tau _*$
such that if $\tau' <\tau _*$ then $\nn{}{\Pi_\rL u(t_3+\tau ' )-
\Pi_\rL u(t_3  )}<\epsilon /4$.
{\em This $\tau _*$ depends only on $\nn{}{\u{\rm f}}$ and $\epsilon $}, as
can be seen from the following argument:
Since $\Pi_\rH \u3=0$, we can choose the
controls in such a way that $\nn{}{\Pi_\rH u(t_3+t )}$ is an
increasing function of $t$ and is therefore bounded by
$\nn{}{\Pi_\rH\u{\rm f}}$. 
The equation for the low-frequency part is then a finite dimensional
ODE in which all high-frequency contributions can be bounded in terms
of $R=\nn{}{\u{\rm f}}$.
\end{likerem}

Combining the estimates we see that
\begin{equs}
\nn{}{u(t_4)-\u{\rm f}}
\,&=\,\nn{}{\Pi_\rL(u(t_4)-\u{\rm f})}+\nn{}{\Pi_\rH(u(t_4)-\u{\rm f})}\\
\,&\le\,\nn{}{\Pi_\rL(u(t_4)-u(t_3))}+\nn{}{\Pi_\rL(u(t_3)-\u{\rm f})}\\
&~~+\nn{}{\Pi_\rH(\u4-\u{\rm f})}~~\le~~\epsilon ~.
\end{equs}
The proof of \theo{theo:control} is complete.
\end{proof}

\section{Strong Feller Property and Proof of \theo{theo:super}}

The aim of this section is to show the strong Feller property of the
process defined by \eref{e:SGL} resp.~\eref{e:SGLL}.
\begin{theorem}
\label{theo:strongfeller}
The Markov semigroup $\CP^t$ defined in \eref{e:PTdef}  is strong Feller.
\end{theorem}

\begin{proof}[of \theo{theo:super}]This proof follows a well-known
strategy, see \eg, \cite{ZDP}.
First of all, there is at least one invariant measure for the process
\eref{e:SGLL}, since for a problem in a finite domain, the semigroup
$t\mapsto e^{-At}$ is compact, and therefore \cite[Theorem 6.1.2]{ZDP} applies.

By the controllability \sref{theo:control}, we deduce, see
\cite[Theorem 7.4.1]{ZDP}, that the
transition probability from any point in $\CH$ to any open set in
$\CH$ cannot vanish, \ie, the Markov process is irreducible. 
Furthermore, by \sref{theo:strongfeller}
the process is strong Feller. 
Therefore we can use Doob's theorem \cite[pp.42--43]{ZDP} to
conclude that the invariant measure is unique.
This completes the proof of \theo{theo:super}.
\end{proof}

Before we start with the proof of \sref{theo:strongfeller}, we explain
our strategy.  
Since the nonlinearity $F$ in \eref{e:SGL} is unbounded, we consider first a
cutoff version $\FR$ 
of $F$:
\begin{equ}
\FR(x)\,=\,\bigl(1-\chi\bigl( \|x\| /(3 \rho)\bigr)\bigr) F(x)\;,
\end{equ}
where $\chi$ is a smooth, non-negative function satisfying 
\begin{equ}
\chi(z)\,=\,\left \{
\begin{array}{rl}
1& \text{if $z>2$,}\\
0& \text{if $z<1$.}
\end{array}
\right .
\end{equ}
Similarly, we define 
\begin{equ}[e:Qrho]
Q_\rho(x)\,=\,Q+ \chi( \|x\| / \rho)\Pi_{\kzero }\;,
\end{equ}
where $\Pi_{\kzero }$ is the projection onto the frequencies below $\kzero $. 
\begin{remark}
These cutoffs have the following effect as a function of $\nn{}{x}$:
\begin{claim}
\item[--]
When $\nn{}{x}\le\rho$ then $Q_\rho(x)=Q$ and $\FR (x)=F(x)$.
\item[--] When  $\rho<\nn{}{x}\le2\rho$ then $Q_\rho (x)$ depends on
$x$ and $\FR(x)=F(x)$.
\item[--] When  $2\rho<\nn{}{x}\le 6\rho$ then all Fourier components
of $Q_\rho(x)$ {\em including the ones below $\kzero$} are non-zero
and $\FR (x)$ is proportional to a $F(x)$ times a factor $\le1$.
\item[--] When  $6\rho<\nn{}{x}$ then all Fourier components
of $Q_\rho(x)$ {\em including the ones below $\kzero$} are non-zero
and $\FR (x)=0$.
\end{claim}
Thus, at high amplitudes, the nonlinearity is truncated to 0 and the
stochastic forcing extends to {\em all} degrees of freedom.
\end{remark}
Instead of \eref{e:SGLL} we then consider the modified problem
\begin{equ}[e:SGLrho]
d\PhiR^t= -A\PhiR^t \,dt + \bigl(\FR\circ\PhiR^t\bigr) \,dt
+ \bigl(Q_\rho\circ\PhiR^t\bigr)\,dW(t)\;,
\end{equ}
with  $\Phi^0_\rho( \ic )= \ic \in\CH$.
Note that the cutoffs are chosen in such a way that the dynamics of
$\PhiR^t( \ic )$ {\em coincides} with that of $\Phi^t( \ic )$ as long as
$\|\Phi^t( \ic )\|<\rho$. Furthermore, as will be seen later, the
definition 
of $Q_\rho$ has been made in such a way as to {\em preserve controllability}.
We will show that the solution of \eref{e:SGLrho} defines a Markov
semigroup 
\begin{equ}
\CPR^t \phi( \ic )=\Exp \bigl(\phi\circ \PhiR^t\bigr)( \ic )\;,
\end{equ}
with the following smoothing property:
\begin{theorem}
\label{theo:main}
There exist exponents $\mu, \nu>0$, and for all $\rho>\rho _0$
there is a constant $C_\rho$
such that for every $\phi \in \CB_b(\CH)$,
for every $t>0$ and for every $ \ic \in\CH$, the function $\CPR^t \phi$ is
differentiable and its 
derivative satisfies
\begin{equ}[e:estDerPt]
\|D \CPR^t \phi( \ic )\| \le C_\rho t^{-\mu} 
(1+\| \ic \|^\nu)
 \|\phi\|_{\L^\infty} \;.
\end{equ}
\end{theorem}

Using this theorem, the proof of \theo{theo:strongfeller} follows from
a limiting argument.

\begin{proof}[of \theo{theo:strongfeller}]
Choose $ x \in \CH$, $t>0$, and $\eps>0$. We denote by $\CB$ the ball of radius
$2\|x\|$ centered around the origin in $\CH$. 
Using \sref{prop:allerlei} we can
find a constant $\rho$ (sufficiently large) such that for every $y \in
\CB$, the 
inequality
\begin{equ}
\Prob\Bigl(\sup_{s \in [0,t]} \bigl\| \Phi^s(y)\bigr\| > \rho\Bigr) \le {\eps
\over 8}
\end{equ}
holds. 
Choose $\phi \in \CB_b(\CH)$ with
$\|\phi\|_{\L^\infty} \le 1$. We have by the triangle inequality
\begin{equs}
\bigl|\CP^t \phi(x) - \CP^t \phi(y)\bigr| &\le\bigl|\CP^t \phi(x) - \CPR ^t
\phi(x)\bigr|  + \bigl|\CPR ^t \phi(x) - \CPR ^t \phi(y)\bigr| \\
&\quad + \bigl|\CP^t \phi(y) - \CPR ^t \phi(y)\bigr| \;.
\end{equs}
Since the dynamics of the cutoff equation
and the dynamics of the original equation coincide on the ball of radius
$\rho$, we can write, for every $z \in \CB$,
\begin{equs}
\bigl|\CP^t \phi(z) - \CPR ^t \phi(z)\bigr| &= \Exp \bigl|\bigl(\phi \circ
\Phi^t\bigr)(z) - \bigl(\phi \circ \PhiR ^t\bigr)(z) \bigr| \\
&\le 2\|\phi\|_{\L^\infty} \,\,\Prob \Bigl(\sup_{s \in [0,t]} \bigl\| \Phi^s(z)\bigr\| >
\rho\Bigr) \le {\eps \over 4}\;.
\end{equs}
This implies that
\begin{equ}
\bigl|\CP^t \phi(x) - \CP^t \phi(y)\bigr| \le  {\eps \over 2}  + \bigl|\CPR ^t
\phi(x) - \CPR ^t \phi(y)\bigr|\;.
\end{equ}
By \theo{theo:main} we see that if $y$ is sufficiently
close to $x$ then
\begin{equ}
\bigl|\CPR^t  \phi(x) - \CPR^t  \phi(y)\bigr| \le {\eps\over 2}\;.
\end{equ}
Since $\epsilon $ is arbitrary we conclude that $\CP^t\phi$ is
continuous when  $\|\phi\|_{\L^\infty} \le 1$. The generalization to any
value of $\|\phi\|_{\L^\infty} $ follows by linearity in $\phi$.
The proof of \theo{theo:strongfeller} is complete.
\end{proof}

\section{Regularity of the Cutoff Process}
\label{sec:reg}

In this section, we start the proof of \theo{theo:main}.
If the cutoff problem were finite dimensional, a result like
\theo{theo:main} could be derived easily using, \eg, the works of H\"ormander \cite{H1,Ho} or
Norris \cite{Norr}. In the present infinite-dimensional context we
need to modify the corresponding techniques, but the general idea retained is Norris'. The main idea will be to treat the
(infinite number of)
high-frequency modes by a method which is an extension of
\cite{ZDP,Cerr}, while the low-frequency part is handled by a variant
of the Malliavin calculus adapted from \cite{Norr}. It is at the
juncture of these two techniques that we need a cutoff in the nonlinearity.

\subsection{Splitting and Interpolation Spaces}
\label{sec:split}

Throughout the remainder of this paper, we will again denote by $\CH_\rL$ and
$\CH_\rH$ the spaces corresponding to the low (resp.~high)-frequency parts. We
slightly change the meaning of ``low-frequency'' by including in the
low-frequency part all those
frequencies that are driven by the noise which are in $\I$ as defined
above. More precisely, the low-frequency part is now 
$\{ k~:~ |k| \le\kmax-1\}$, where
$\kmax=\max\{k~:~k\in\I\}+1$.
The important fact is that
\begin{claim} 
\item[--] $\CH_\rL$ is finite dimensional and
\item[--] for every unforced frequency $k$, there exist three different
forced frequencies
$k_1,k_2,k_3$ in the low-frequency part for which $k_1 + k_2 + k_3 =
k$, and, as explained above, none of these frequencies is in the
triplet for another $k$.
\end{claim}
Since $A = 1-\Delta$ is diagonal with respect to this splitting, we can define
its low (resp.~high)-frequency parts $A_\rL$ and $A_\rH$ as operators on
$\CH_\rL$ and $\CH_\rH$. From now on, $L$ will always denote the
dimension of $\CH_\rL$, which will therefore be identified with
$\R^L$.\footnote{The choice of $L$ above is dictated by the desire to
obtain a dimension equal to $L$ and not $L+1$.} 
We also allow ourselves to switch freely between equivalent norms on
$\R^L$, when deriving the various bounds.

In the sequel, we will always use the notations $D_\rL$ and $D_\rH$ to denote
the derivatives with respect to $\CH_\rL$ (resp.~$\CH_\rH$) of a differentiable
function defined on $\CH$. The words ``derivative'' and ``differentiable'' will
always be understood in the strong sense, \ie, if $f:\SB_1 \to \SB_2$ with
$\SB_{1}$ and $\SB_{2}$ some Banach spaces, then $Df : \SB_1 \to
\SL(\SB_1,\SB_2)$, \ie, it is bounded from $\SB_1$ to $\SB_2$.

We introduce the interpolation spaces $\SH^\gamma$ (for every $\gamma \ge
0$) defined as being equal to the domain of $A^\gamma$ equipped with the graph
norm
\begin{equ}
\|x\|^2_\gamma =\|A^{\gamma} x\|^2=\|(1-\Delta)^{\gamma} x\|^2\;.
\end{equ}
Clearly, the $\CH^\gamma$ are Hilbert spaces and we have the inclusions
\begin{equ}
\SH^\gamma \subset \SH^\delta \quad\text{if}\quad \gamma \ge \delta\;.
\end{equ}
Note that in usual conventions, $\CH^\gamma $ would be the Sobolev
space of index $2\gamma +1$. Our motivation for using non-standard
notation comes from the fact that our basic space is that with {\em
one} derivative, which we call $\CH$, and that $\gamma $ measures
additional smoothness in terms of powers of the generator of the
linear part.


\subsection{Proof of \theo{theo:main}}

The proof of \theo{theo:main} is based on \sref{prop:reg}
and \sref{prop:high} which we now state.

\begin{proposition}
\label{prop:reg}
Assume that the noise satisfies condition \eref{e:coeff}. Then \eref{e:SGLrho}
defines a stochastic flow $\PhiR^t $ on $\CH$ with the following
properties which hold for any $p\ge1$:
\begin{claim}
%
%
\item[\myitem{A}] If $ \ic \in \SH^\gamma$ with some $\gamma$ satisfying 
$0 \le \gamma
\le \alpha$, the solution of \eref{e:SGLrho} stays in
$\SH^\gamma$, with a bound
\minilab{e:boundPhi}
\begin{equ}[e:boundPhi2]
\Exp\Bigl(\;\sup_{0<t<T} \|\PhiR^t ( \ic )\|_\gamma^p \Bigr) \le C_{T,p,\rho}
(1+\| \ic \|_\gamma)^p\;.
\end{equ}
If $\gamma\ge 1$ the solution exists in the strong sense in $\CH$.
%
%
\item[\myitem{B}] The quantity $\PhiR^t ( \ic )$ is in $\SH^\alpha$ with probability $1$ for every time
$t>0$ and every $ \ic \in \CH$. Furthermore, for every
$T>0$ there is a constant $C_{T,p,\rho } $ for which
\minilab{e:boundPhi}
\begin{equ}[e:boundPhi1]
\Exp\Bigl(\;\sup_{0<t<T} t^{\alpha p }\|\PhiR^t ( \ic )\|_\alpha^p
\Bigr) \le C_{T,p,\rho }  
(1+\| \ic \|)^{p}\;.
\end{equ}
%
%
\item[\myitem{C}] The mapping $\xi \mapsto \PhiR^t(\xi)$ (for $\omega$
and $t$ fixed) has {a.s.} bounded partial derivatives with respect to
$\xi$. Furthermore, we have for every $\xi, h \in \CH$ the bound
\minilab{e:boundPhi}
\begin{equ}[e:boundDPhi1]
\Exp\Bigl(\;\sup_{0<t<T} \bigl\|\bigl(D\PhiR^t ( \ic )\bigr)h\bigr\|^p
\Bigr) \le C_{T,p,\rho}\|h\|^p\;,
\end{equ}
for every $T>0$.
%
%
\item[\myitem{D}] For every $h \in \CH$ and $ \ic \in \CH^\alpha$, the
quantity $\bigl(D\PhiR^t ( \ic )\bigr)h$ is in $\SH^\alpha$ with
probability $1$ for every $t>0$. Furthermore, for a $\nu$
depending only on $\alpha $ the bound
\minilab{e:boundPhi}
\begin{equ}[e:boundDPhi2]
\Exp\Bigl(\;\sup_{0<t<T} t^{\alpha  p}\bigl\|\bigl(D\PhiR^t ( \ic )\bigr)h\bigr\|_\alpha^p
\Bigr) \le C_{T,p,\rho}(1+\| \ic \|_\alpha )^{\nu p}\|h\|^p\;,
\end{equ}
holds for every $T>0$.
\item[\myitem{E}] For every $ \ic \in \CH^\gamma$ with $\gamma \le \alpha$, we have the small-time estimate
\minilab{e:boundPhi}
\begin{equ}[e:boundPhi3]
\Exp\Bigl(\;\sup_{0<t<\eps} \nn[b]{\gamma}{\PhiR^t ( \ic ) - e^{-At} \xi}^p \Bigr) \le C_{T,p,\rho}
(1+\| \ic \|_\gamma)^p \eps^{p/16}\;,
\end{equ}
which holds for every $\eps \in (0,T]$ and every $T>0$.
\end{claim}
\end{proposition}
This proposition will be proved in \sect{sec:propreg}.
\begin{proposition}
\label{prop:high}
There exist a time $T^* > 0$ and exponents $\mu,\nu>0$ such that for every
$\phi \in \CC^2_b(\CH)$, every $ \ic\in\SH^\alpha$ and every $t\le T^*$,
\begin{equ}[e:ehigh]
\|D\CPR^t \phi( \ic )\| \le C t^{-\mu} \bigl(1+\| \ic \|_\alpha^\nu \bigr)
\|\phi\|_{\L^\infty}\;.
\end{equ}
\end{proposition}

\begin{proof}[of \theo{theo:main}]
We first prove the bound for the case
$\phi \in \CC_b^2(\CH)$. Let $h\in\CH$.
Using the definition \eref{e:PTdef} of $\CPR^t \phi$
and the Markov property of the flow we write
\begin{equs}
\|D\CPR^{2t}\phi( \ic )h\| &= \bigl\|D \Exp \bigl(\CPR^t   \phi \circ
\PhiR^t \bigr)( \ic )h\bigr\|
= \Bigl\|\Exp \Bigl(\bigl(D \CPR^t   \phi \circ
\PhiR^t \bigr)( \ic )D\PhiR^t ( \ic )h\Bigr)\Bigr\| \\
&\le \sqrt{\Exp \bigl\|\bigl(D \CPR^t   \phi \circ
\PhiR^t \bigr)( \ic )\bigr\|^2}\sqrt{\Exp \bigl\|D\PhiR^t
( \ic )h\bigr\|^2} \;.
\end{equs}
Bounding the first square root by \sref{prop:high} and then applying
\sref{prop:reg}, we get a bound
\begin{equs}
\|D\CPR^{2t}\phi( \ic )h\|&\le C \|\phi\|_{\L^\infty} t^{-\mu}\sqrt{\Exp \bigl(1 +
\|\PhiR^t ( \ic )\|_\alpha^\nu\bigr)^2} \sqrt{\Exp \bigl\|D\PhiR^t
( \ic )h\bigr\|^2}\\
&\le C\|\phi\|_{\L^\infty}  t^{-\mu}t^{-\alpha \nu}(1+\| \ic \|)^{\nu}\|h\|
\;.
\end{equs}
Thus, we have shown \eref{e:estDerPt} when $\phi \in \CC_b^2(\CH)$.
The method of extension to
arbitrary $\phi \in \CB_b(\CH)$ can be found in \cite[Lemma 7.1.5]{ZDP}.
The proof of \theo{theo:main} is complete.
\end{proof}

\subsection{Smoothing Properties of the Transition Semigroup}

In this subsection we prove the smoothing bound \sref{prop:high}.
Thus,
we will no longer be
interested in smoothing in position space as shown in \sref{prop:reg}
but in smoothing properties of the 
transition semigroup associated to \eref{e:SGLrho}.

\begin{likerem}[Important remark.]
In this section and up to \sect{sec:theend} we always
tacitly assume that we are considering the cutoff equation
\eref{e:SGLrho} and we will omit the index $\rho $.
\end{likerem}

Thus, we will write 
Eq.\eref{e:SGLrho} as
\begin{equ}[e:cutoff]
d\Phi^t      = -A \Phi^t\,dt + \bigl(F \circ
\Phi^t \bigr)\,dt  + \bigl(Q \cp\bigr )\, dW(t)\;.
\end{equ}

The solution of \eref{e:cutoff}
generates a semigroup on the space $\CB_b(\CH)$ of bounded
Borel functions over $\CH = \CH_{\rL} \oplus \CH_{\rH}$ by
\begin{equ}
\CP^t \phi = \Exp\bigl(\phi \circ \Phi^t\bigr)\;,\qquad
\phi\in \CB_b(\CH)\;.
\end{equ}
Our goal will be to show that the mixing properties of the nonlinearity are
strong enough to make $\CP^t \phi$ differentiable, even if $\phi$ is only
measurable.

We will need a separate treatment of the high and low frequencies, and
so we
reformulate \eref{e:cutoff} as
\minilab{e:SDE}
\begin{equs}[2]
d\Phi^t_{\rL} &= -A_{\rL} \Phi^t_{\rL}\,dt + \bigl(F_{\rL}\circ
\Phi^t\bigr)\,dt + \bigl(Q_\rL\cp\bigr )\, dW_{\rL}(t)\;,\quad&\quad \Phi^t_{\rL} &\in
\CH_{\rL}\;,\quad
\label{e:SDE1}\\
d\Phi^t_{\rH} &= -A_{\rH} \Phi^t_{\rH}\, dt + \bigl(F_{\rH}\circ
\Phi^t\bigr)\,dt + Q_\rH \, dW_{\rH}(t)\;,\quad&\quad \Phi^t_{\rH} &\in
\CH_{\rH}\;,\label{e:SDE2}
\end{equs}
where $\CH_{\rL}$ and $\CH_\rH$ are defined in
Section~\ref{sec:split} and the cutoff version of $Q$ was defined in
\eref{e:Qrho}. Note that $Q_\rH\bigl(\Phi^t( \ic )\bigr)$ is independent
of $ \ic $ and $t$ by construction, which is why we can use $Q_\rH$ in 
\eref{e:SDE2}.

The proof of \sref{prop:high} is based on the following two results
dealing with the low-frequency part and the cross-terms between low
and high frequencies, respectively.
\begin{proposition}
\label{prop:lowone}
There exist a time $T^*>0$ and exponents $\mu,\nu>0$ such that for every
$\phi \in \CC^2_b(\CH)$, every $ \ic \in\SH^\alpha$ and every~ $t\le T^*$,
\begin{equ}
\Bigl\|\Exp \Bigl(\bigl(D_{\rL}\phi\circ\Phi^t\bigr)( \ic )(D_\rL
\Phi_\rL^t)( \ic )\Bigr)\Bigr\| \le C t^{-\mu} \bigl(1+\| \ic \|_\alpha^\nu\bigr)
\|\phi\|_{\L^\infty} \;.
\end{equ}
\end{proposition}
\begin{lemma}
\label{lem:cross}
For every $T>0$ and every $p\ge 1$, there are constants $C_1, C_2 > 0$ such
that for every $t\le T$, one has the estimates (valid for
$h_\rL\in\CH_\rL$ and $h_\rH\in\CH_\rH$):
\minilab{e:crossbound}
\begin{equs}
\Exp \sup_{0<s<t} \bigl\|\bigl(D_\rL \Phi^s_\rH\bigr)( \ic )h_\rL\bigr\|^p &\le
C_1t^p\nn{}{h_\rL}^p\;,\label{e:cross1}\\
\Exp \sup_{0<s<t} \bigl\|\bigl(D_\rH \Phi^s_\rL\bigr)( \ic )h_\rH\bigr\|^p &\le C_2
 t^{p/4}\nn{}{h_\rH}^p\;.\label{e:cross2}
\end{equs}
These bounds are independent of $ \ic \in\CH$.
\end{lemma}
\begin{remark}In the absence of the cutoff $\rho$
one can prove inequalities
like
\eref{e:crossbound}, but with an additional factor of $(1+\| \ic \|^2
)^p$ on the right. This is not good enough for our strategy and is
the reason for introducing a cutoff.
\end{remark}

The proof of \sref{prop:lowone} will be given in \sect{sec:malliavin}
and the proof of \lem{lem:cross} will be given in \sect{sec:cross}.
\begin{proof}[of \prop{prop:high}]
The proof will be performed in the spirit of \cite{ZDP} and \cite{Cerr}, using
a modified version of the Bismut-Elworthy formula.
Take a function $\phi \in \CC^2_b(\CH)$. 
We consider $Q_{\rL}$ and $Q_{\rH}$ as acting on and into
$\CH_{\rL}$ and
$\CH_{\rH}$ respectively. It is possible to write as a consequence of It\^o's
formula:
\begin{equs}
\bigl(\phi \circ \Phi^t\bigr)( \ic ) &= \CP^t \phi( \ic ) + \int_0^t
\bigl((D\CP^{t-s}\phi)\circ \Phi^s\bigr)( \ic )\,\bigl(Q\circ \Phi^s\bigr )( \ic )\,dW(s) \\
	&= \CP^t \phi( \ic ) + \int_0^t
\bigl((D_{\rL}\CP^{t-s}\phi)\circ
\Phi^s\bigr)( \ic )\,\bigl(Q_{\rL}\circ\Phi^s\bigr )( \ic )\,dW_\rL(s)
\\
	&\qquad + \int_0^t \bigl((D_{\rH}\CP^{t-s}\phi)\circ
\Phi^s\bigr)( \ic )\,Q_{\rH}\,dW_\rH(s)\;.\label{e:evolphi}
\end{equs}
Choose some $h \in \CH_\rH$.
By \sref{prop:reg}, $\bigl(D_{\rH} \Phi^t_{\rH}\bigr)( \ic )h$ is in $ \SH^\alpha$ for
positive times and is bounded by \eref{e:boundDPhi2}. Using condition
\eref{e:zzz2} we see that $Q_\rH^{-1}$ maps to $\CH_\rH$
and so we can multiply both sides of \eref{e:evolphi} by
\begin{equ}
\int_{t/4}^{3t/4} \scal[B]{Q_{\rH}^{-1} \bigl(D_{\rH}
\Phi^s_{\rH}\bigr)( \ic )h\,,\, dW_\rH(s)}\;,
\end{equ}
where the scalar product is taken in $\CH_\rH$.
Taking expectations on both sides, the first two terms on the right
vanish
because $dW_\rL$ and $dW_\rH$ are independent and of mean zero.
Thus, we get
\begin{equa}[e:expSemi]
{}&\Exp\Bigl( \bigl(\phi \circ \Phi^t\bigr)( \ic )\int_{t/4}^{3t/4}
\scal[B]{Q_{\rH}^{-1} \bigl(D_{\rH}
\Phi_\rH^s\bigr)( \ic )h\,,\, dW_\rH(s)} \Bigr) \\
	&\qquad\qquad = \Exp \int_{t/4}^{3t/4}
{\bigl((D_{\rH}\CP^{t-s}\phi)\circ \Phi^s\bigr)( \ic )
\,\bigl(D_{\rH}
\Phi_\rH^s\bigr)( \ic )h}\,ds
\;,
\end{equa}
We add to both sides of \eref{e:expSemi} the term
\begin{equ}
\Exp \int_{t/4}^{3t/4} 
\bigl((D_{\rL}\CP^{t-s}\phi)\circ
\Phi^s\bigr)( \ic )\,\bigl(D_{\rH}
\Phi^s_\rL\bigr)( \ic )h\,ds\;,
\end{equ}
and note that the r.h.s.~can be rewritten as
\begin{equ}
{\int_{t/4}^{3t/4}D_{\rH}\Exp 
\bigl((\CP^{t-s}\phi)\circ \Phi^s\bigr)( \ic )h\,ds} 
= {t\over 2}
{D_{\rH}\Exp 
\bigl(\phi\circ \Phi^t\bigr)( \ic )h}\;,
\end{equ}
since by the Markov property,
$\Exp\bigl(\CP^{t-s}\phi\circ \Phi^s\bigr )( \ic )=\Exp\bigl(\phi\circ
\Phi^t\bigr )( \ic )$. 
Therefore, \eref{e:expSemi} leads to
\begin{equa}[e:estHigh]
\bigl(D_{\rH}\CP^t \phi\bigr)( \ic )h  &= {2 \over t}\Exp\Bigl(
\bigl(\phi \circ\Phi^t\bigr)( \ic )\int_{t/4}^{3t/4} \scal[b]{Q_{\rH}^{-1}
\bigl(D_{\rH}
\Phi_\rH^s\bigr)( \ic )h, dW_\rH(s)} \Bigr)\qquad \\
&\quad + {2 \over t}\Exp \int_{t/4}^{3t/4} \bigl((D_{\rL}\CP^{t-s}\phi)\circ
\Phi^s\bigr)( \ic )\,\bigl(D_{\rH}
\Phi^s_\rL\bigr)( \ic )h\,ds \;.
\end{equa}
For the low-frequency part, we use the equality
\begin{equa}[e:estLow]
\bigl(D_\rL \CP^t \phi\bigr)( \ic ) &= \Exp \Bigl(\bigl(D_\rL \CP^{t/2}\phi \circ
\Phi^{t/2}\bigr)( \ic ) \bigl(D_\rL \Phi_\rL^{t/2}\bigr)( \ic )\Bigr) \\
&\qquad +  \Exp \Bigl(\bigl(D_\rH \CP^{t/2}\phi \circ \Phi^{t/2}\bigr)( \ic )
\bigl(D_\rL \Phi_\rH^{t/2}\bigr)( \ic )\Bigr)\;.
\end{equa}
We introduce the Banach spaces $\SB_{T,\mu_*,\nu_*}$ of measurable
functions $f : 
(0,T)\times \SH^\alpha \to \CH$, for which
\begin{equ}[e:Tnorm]
\norm f\norm _{T,\mu_*,\nu_*} \equiv \sup_{0<t<T} \;\sup_{ \ic \in
\SH^\alpha} 
{t^{\mu_*} \|f(t, \ic )\| \over
1+\| \ic \|_\alpha^{\nu_*}}
\end{equ}
is finite.
Choose $\mu_*$ as the maximum of the constants $\alpha $ and the $\mu$
appearing in \sref{prop:lowone}.
Similarly $\nu_*$ is the maximum of the $\nu$ of
\sref{prop:reg}~\myitem{D} and
the one in \sref{prop:lowone}.

We will show that there exists a $T>0$ such that $f_\phi :(t, \ic )\mapsto
\bigl(D\CP^t \phi\bigr)( \ic )$ belongs to $\SB_{T,\mu_*,\nu_*}$ and that $\norm f_\phi\norm _{T,\mu_*,\nu_*} \le
C\|\phi\|_{\L^\infty}$, thus proving  \sref{prop:high}. The fact that $f_\phi \in \SB_{T,\mu_*,\nu_*}$ for
every $T$ if $\phi \in \CC_b^2(\CH)$ is shown in \cite{ZDP}, so we only have to
show the bound on its norm.

By \sref{prop:reg}, Eqs.\eref{e:cross2} and \eref{e:estHigh},
and the Cauchy-Schwarz inequality, we have the estimate, valid for
$h\in\CH_\rH$:
\begin{equs}
{}\bigl|\bigl(D_{\rH}&\CP^t \phi\bigr)( \ic )h\bigr|  
\le \|\phi\|_{\L^\infty} 
{2\over t}\left(\Exp \int_{t/4}^{3t/4} \bigl\|Q_{\rH}^{-1} \bigl(D_{\rH}
\Phi_\rH^s\bigr)( \ic )h\bigr\|^2\,ds\right)^{1/2} \\
&\quad + {2\over t} \norm f_\phi\norm _{t,\mu_*,\nu_*} \,\Exp\int_{t/4}^{3t/4}
{1+\|\Phi^s( \ic )\|_\alpha^{\nu_*}\over (t-s)^{{\mu_*}}} \bigl\|\bigl(D_\rH
\Phi_\rL^s\bigr)( \ic )h \bigr\|\,ds \\
&\le C t^{-\alpha  } \|\phi\|_{\L^\infty}\bigl(1+\nn{\alpha
}{ \ic }^{\nu_*}\bigr)
\|h\| \\&\quad + Ct^{-\mu_*} 
\norm f_\phi\norm _{t,\mu_*,\nu_*}
\left({\Exp \sup_{s \in [{t\over 4},{3t\over 4}]}
\bigl(1+\|\Phi^s( \ic )\|_\alpha^{\nu_*}\bigr)^2}\right)^{1/2} 
\\&\qquad\times
 \left(\Exp \sup_{s \in [{t\over 4},{3t\over 4}]}
\bigl\|\bigl(D_\rH \Phi^s_\rL\bigr)( \ic )h \bigr\|^2\right)^{1/2}
\\
& \le C t^{-\alpha  } \|\phi\|_{\L^\infty}\bigl(1+\nn{\alpha
}{ \ic }^{\nu_*}\bigr) \|h\|+ C t^{-\mu_*+1/4} \norm f_\phi\norm _{t,\mu_*,\nu_*}
(1+\| \ic \|_\alpha^{\nu_*})\|h\|\;.
\end{equs}
For the low-frequency part, we use \sref{prop:lowone},
Eqs.\eref{e:cross1} and \eref{e:estLow}, and the property that
$\|\CP^t\phi\|_{\L^\infty} \le \|\phi\|_{\L^\infty}$ and find for
$h\in\CH_\rL$,  
\begin{equs}
\bigl|\bigl(&D_{\rL}\CP^t \phi\bigr)( \ic )h\bigr|  \le C
t^{-{\mu_*}}\|\phi\|_{\L^\infty} 
\bigl(1+\| \ic \|_\alpha^{\nu_*}\bigr)\,\nn{}{h} \\
&\quad + C t^{-\mu_*}\norm f_\phi\norm _{t,\mu_*,\nu_*} \Exp\Bigl(\bigl(
{1+\|\Phi^{t/2}( \ic )\|_\alpha^{\nu_*}}\bigr) \bigl\|\bigl(D_\rL
\Phi_\rH^{t/2}\bigr)( \ic )h\bigr\|\Bigr)
\\[2mm]
&\le Ct^{-{\mu_*}}\|\phi\|_{\L^\infty} \bigl(1+\| \ic
\|_\alpha^{\nu_*}\bigr)\,\nn{}{h} \\ 
&\quad + C t^{-\mu_*}\norm f_\phi\norm _{t,\mu_*,\nu_*} \sqrt{\Exp \bigl(
{1+\|\Phi^{t/2}( \ic )\|_\alpha^{\nu_*}}\bigr)^2}\sqrt{\Exp \bigl\|\bigl(D_\rL
\Phi_\rH^{t/2}\bigr)( \ic )h\bigr\|^2} 
\\[2mm]
&\le Ct^{-{\mu_*}}\|\phi\|_{\L^\infty} \bigl(1+\| \ic \|_\alpha^{\nu_*}\bigr) \,\nn{}{h}+
C
t^{-\mu_*+1}\norm f_\phi\norm _{t,\mu_*,\nu_*} \bigl(
{1+\| \ic \|_\alpha^{\nu_*}}\bigr)\,\nn{}{h} \;. 
\end{equs}
Combining the above expressions we get for every $T<T^*$ a bound
of the type
\begin{equ}
\norm  f_\phi\norm _{T,\mu_*,\nu_*} \le C_1 \|\phi\|_{\L^\infty} + C_2 T^{1/4} \norm  f_\phi\norm _{T,\mu_*,\nu_*}\;.
\end{equ}
Choosing $T^{1/4} < 1/(2C_2)$, we find
\begin{equ}[e:endlich]
\norm  f_\phi\norm _{T,\mu_*,\nu_*} \le
C\|\phi\|_{\L^\infty}\;.
\end{equ}
Since $f_\phi(t, \ic )=\bigl(D \CP^t\phi\bigr)( \ic )$, inspection of
\eref{e:Tnorm}
shows that \eref{e:endlich} is equivalent to \eref{e:ehigh}. The proof
of \sref{prop:high} is complete. 
\end{proof}

\section{Malliavin Calculus}
\label{sec:malliavin}

To prove  \sref{prop:lowone} we will apply a modification of
Norris'
version
of the Malliavin calculus. This modification takes into account some
new features which are necessary due to our splitting of the problem
in high and low frequencies (which in turn was done to deal with the
infinite dimensional nature of the problem).
 
Consider first the deterministic
PDE for a flow:
\begin{equ}[deterministic]
{d\Psi^t( \ic )\over dt} = -A\Psi^t( \ic ) +\bigl( F\circ \Psi^t\bigr )( \ic )\;.
\end{equ}
This is really an abstract reformulation for the flow defined by the
GL equation, and $ \ic $ belongs to a space $\CH$, which for our problem
is a suitable Sobolev space. The linear operator $A$ is chosen as
$1-\Delta$, while the non-linear term $F$ corresponds to $2u-u^3$ in
the GL equation. Below, we will work with approximations to the
GL equation, and all we need to know is that $A:\CH\to\CH$ is the
generator of a strongly continuous semigroup, and $F$ will be seen to
be bounded with bounded derivatives.

For each fixed $ \ic \in\CH$ we consider the
following stochastic variant of \eref{deterministic}:
\begin{equ}[stochastic]
d\Psi^t( \ic ) = -A\Psi^t( \ic )\,dt +\bigl( F\circ \Psi^t\bigr )( \ic )\,dt +
\bigl(Q\circ \Psi^t)( \ic )\,dW(t)\;.
\end{equ}
with initial condition $\Psi^0( \ic ) =  \ic $.
Furthermore, $W$ is the cylindrical
Wiener process on a separable Hilbert space $\CW$ and
$Q$ is a strongly differentiable map from $\CH$ to $\cscr L^2(\CW,\CH)$,
the space of bounded linear Hilbert-Schmidt operators from $\CW$ to $\CH$.

We next introduce the notion of directional derivative (in the
direction of the noise) and the reader
familiar with this concept can pass directly to \eref{e:defDY}.
To understand this concept consider first the case of a function
$t\mapsto v^t_i\in\CW$. Then the variation $\SD _{v_i}\Psi^t$ of $\Psi^t$
in the direction
$v_i$ is obtained by replacing $dW(t)$ by $dW(t)+\epsilon v_i^t\,dt$
and
it satisfies the equation
\begin{equs}
d \SD_{v_i} \Psi^t\,&=\,\bigl(-A\SD _{v_i} \Psi^t + (DF\circ
\Psi^t)\SD _{v_i} \Psi^t\bigr )\,dt+\bigl((DQ\circ\Psi^t)\SD _{v_i} \Psi^t\bigr
)\,dW(t) \\&~~+(Q\circ\Psi^t) v_i^t\,dt\;.
\end{equs}
Intuitively, the first line comes from varying $\Psi^t$ with respect
to the noise and the second comes from varying the noise itself.

We will need
a finite number $L$ of directional derivatives, and so we introduce some
more general notation. We combine $L$ vectors $v_i$ as used above
into a matrix called $v$ which is an element of
$\Omega \times [0,\infty) \to\CW^L$.
We identify $\CW^L$ with $\cscr L(\R^L,\CW)$.
Note that we now allow $v$ to depend on $\Omega$, and to make things
work, we require $v$ to be a predictable stochastic process, \ie, $v^t$
only depends on the noise before time $t$.
The stochastic process $G_v^t \in \CH^L$ (corresponding to $\SD
_v\Psi^t$)
is then defined as the solution of the equation
\begin{equa}[e:defDY]
dG^t_v h &= \Bigl(-AG^t_v  +
\bigl(DF\circ\Psi^t\bigr )G^t_v + \bigl(Q\circ\Psi^t\bigr ) v^t \Bigr)h \,dt
\\
	&\qquad + \Bigl(\bigl(DQ\circ \Psi^t\bigr)\,G^t_v h\Bigr)\,dW(t)\;,\\
\quad G^0_v &= 0 \;,
\end{equa}
which has to hold for all $h\in\R^L$.

Having given the detailed definition of $G^t_v$, we will denote it
henceforth by the more suggestive
$$
G^t_v ( \ic )\,=\,\SD_v \Psi^t( \ic)\;,
$$
to make clear that it is a directional derivative.
We use the notation $\SD_v $ to distinguish this derivative from the
derivative $D$ with respect to the initial condition $ \ic $.

For \eref{stochastic} and
\eref{e:defDY} to make sense, two assumptions on $F$, $Q$ and
$v$ are needed:
\begin{claim}
\item[\bf A1] $F:\CH \to \CH$ and $Q : \CH \to \SL^2(\CW,\CH)$ are of
at most linear growth and have bounded first and second derivatives. 
\item[\bf A2] The stochastic process $t \mapsto v^t$ is predictable,
has a continuous version, and satisfies 
\begin{equ}
\Exp \Bigl(\sup_{s \in [0,t]} \|v^s\|^p \Bigr)< \infty\;,
\end{equ}
for every $t>0$ and every $p\ge 1$. (The norm being the norm of $\CW^L$.)
\end{claim}
It is easy to see that these conditions imply the hypotheses of
\theo{theo:exist} for the problems \eref{stochastic} and
\eref{e:defDY}. Therefore 
$G^t_v$ is a well-defined strongly Markovian
stochastic process. 

With these notations one has the well-known Bismut integration by parts
formula \cite{Norr}.

\begin{proposition}
\label{prop:parts}
Let $\Psi^t$ and $\SD_v \Psi^t$ be defined as above and assume {\bf A1} and {\bf A2} are satisfied.
Let $\CB \subset \CH$ be an open subset of $\CH$ such that $\Psi^t \in \CB$ almost surely and let $\phi : \CB \to \R$ be a differentiable function such that
\begin{equ}
\Exp \|\phi(\Psi^t)\|^2 + \Exp \|D\phi(\Psi^t)\|^2 < \infty\;.
\end{equ}
Then we have for every $h \in \R^L$ the following
identity in $\R$:
\begin{equ}[e:parts]
\Exp\bigl(D\phi(\Psi^t)\SD_v \Psi^t h \bigr)
= \Exp\biggl(\phi(\Psi^t)\int_0^t
\scal[b]{v^s h,dW(s)}\biggr)\;,
\end{equ}
where $\scal{\cdot,\cdot}$ is the scalar product of\/ $\CW$.
\end{proposition}
\begin{remark}The Eq.\eref{e:parts} is useful because it relates the
expectation of $D\phi$ to that of $\phi$. In order to fully exploit
\eref{e:parts} we will need to get rid of the factor $\SD_v
\Psi^t$. This will be possible by a clever choice of $v$.
This procedure is explained for example in \cite{Norr} but we will
need a new variant of his results because of the high-frequency part. In the
sequel, we will proceed in two steps. 
{\em We need only bounds on $D_\rL\phi$,
since the smoothness of the high-frequency part follows by other
means.} Thus, it {\em suffices} to construct $\SD_v \Psi^t$ in such a
way that
$\Pi_\rL \SD_v \Psi^t$ is invertible, where $\Pi_\rL$ is the
orthogonal projection onto $\CH_\rL$. The construction of $\Pi_\rL
\SD_v \Psi^t$ follows closely the presentation of
\cite{Norr}. However, we also want $\Pi_\rH \SD_v \Psi^t=0$ and
this elimination of the high-frequency part seems to be new.
\end{remark}

\begin{proof}
The finite dimensional case is stated (with slightly different assumptions on
$F$)
in \cite{Norr}. The extension to the infinite-dimensional setting can be done
without major difficulty. By {\bf A1}--{\bf A2} and \sref{theo:exist},
we ensure that all the expressions appearing in the proof and the
statement are well-defined. By {\bf A2}, we can use It\^o's formula to
ensure the validity of the assumptions for the infinite-dimensional
version of Girsanov's theorem \cite{ZDP}. 
\end{proof}

\subsection{The Construction of $v$}
In order to use
\sref{prop:parts} we will construct $v =
(v_{\rL},v_{\rH})$ in such a way that the high-frequency part of $\SD_v
\PT{t}= (\SD_v
\PT{t}_{\rL},\SD_v \PT{t}_{\rH})$ vanishes. 
This construction is new and will be explained in detail in this subsection.

\begin{likerem}[Notation.]
The equations which follow are quite involved. To keep the notation at
a reasonable level without sacrificing precision we will adopt the
following conventions:
\begin{equs}
(D_\rL F_\rL)^t \,&\equiv\,(D_\rL F_\rL)\circ\Phi^t\;,\\
(D_\rL Q_\rL)^t \,&\equiv\,(D_\rL Q_\rL)\circ\Phi^t\;,
\end{equs}
and similarly for other derivatives of the $Q$ and the $F$.
Furthermore, the reader should note that $D_\rL Q_\rL$ is a linear map
from
$\CH_\rL$ to the linear maps $\CH_\rL\to\CH_\rL$ and therefore, below,
$(D_\rL Q_\rL)h$ with $h\in\CH_\rL$ is a linear map
$\CH_\rL\to\CH_\rL$.
The dimension of $\CH_\rL$ is $L<\infty $.
\end{likerem}
Inspired by \cite{Norr}, we define
the $L\times L$ matrix-valued stochastic processes $U_\rL^t$ and $V_\rL^t$ by
the following SDE's, which must hold for every $h\in\CH_\rL$:
\minilab{e:Jac}
\begin{equs}
dU_\rL^t h &= -A_\rL  U_\rL^t h\,dt + (D_\rL  F_\rL )^t U_\rL^t h\,dt
+\bigl((D_\rL Q_\rL)^t U_\rL^th\bigr)\,dW_\rL(t)\;,
 \\
U_\rL^0 &= I \in  \SL(\CH_{\rL},\CH_\rL) \;,\label{e:Jac1}\\[4mm]
dV_\rL^t h&= V_\rL^t A_\rL h\,dt - V_\rL^t (D_\rL  F_\rL )^t h\,dt- 
V_\rL^t \bigl((D_\rL Q_\rL)^t h\bigr)\,dW_\rL(t)\\
&\qquad + \sumiL V^t_\rL \Bigl( (D_\rL Q_\rL)^t \bigl((D_\rL
Q_\rL)^t h\bigr)e_i\Bigr)e_i \,dt\;,\\ 
V_\rL^0 &= I \in  \SL(\CH_{\rL},\CH_\rL) \;.\label{e:Jac2}
\end{equs}
The last term in the definition of $V_\rL^t$ will be written as
\begin{equ}
\sumiL V^t_\rL \bigl((D_\rL Q^i_\rL)^t \bigr)^2 h \,dt\;,
\end{equ}
where $Q_\rL^i$ is the $i^{\rm th}$ column of the matrix $Q_\rL$.

For small times, the process $U_\rL ^t$ is an approximation to the partial
Jacobian $D_\rL \PT{t}_\rL$, and $V_\rL^t$ is an approximation to its inverse.

We first make sure that the objects in \eref{e:Jac} are well-defined.
The following lemma summarizes the properties of $U_\rL$ and $V_\rL$
which we need later.
\begin{lemma}
\label{lem:VL}
The processes $U_\rL^t$ and $V_\rL^t$ satisfy the following
bounds. For every $p\ge1$ and all $T>0$ there
is a constant $C_{T,p,\rho}$ independent of the initial data (for
$\Phi^t$)
such that 
\minilab{e:VL}
\begin{equs}
\Exp \sup_{t\in[0,T]} \bigl(\nn{}{U_\rL^t}^p+\nn{}{V_\rL^t}^p\bigr)&\le
C_{T,p,\rho}\;,\label{e:VLa}\\
\Exp \bigl(\sup_{t\in[0,\epsilon ]} \nn{}{V_\rL^t-I}^p\bigr)&\le
C_{T,p,\rho}\epsilon ^{p/2}\;,\label{e:VLb}
\end{equs}
for all $\epsilon <T$.
Furthermore, $V_\rL$ is the inverse of $U_\rL$ in the sense that 
$V_\rL^{t} = (U_\rL^t)^{-1}$ almost surely %
\end{lemma}
\begin{proof}The bound \eref{e:VLa} is a straightforward application
of \sref{theo:exist} whose conditions are easily checked. (Note that
we are here in a finite-dimensional, linear setting.)
To prove \eref{e:VLb}, note that $I$ is the initial condition for
$V_\rL$. One writes \eref{e:Jac2} in its integral form and then the
result follows by applying \eref{e:VLa}. 
The last statement can be shown easily by applying It\^o's formula to
the product
$V_\rL^{t}U_\rL^t$. (In fact, the definition of $V_\rL$ was precisely
made with this in mind.)
\end{proof}

We continue with the construction of $v$.
Since $A$ and $Q$ are diagonal with respect to the splitting $\CH = \CH_{\rL}
\oplus \CH_{\rH}$, we can write
\eref{e:defDY} as
\minilab{e:DX}
\begin{equs}
d\,\SD_v \PT{t}_{\rL} &= \Bigl(-A_{\rL}\,\SD_v \PT{t}_{\rL} +
(D_{\rL}F_{\rL})^t \,\SD_v \PT{t}_{\rL} \label{e:DXL}
\\
	& \qquad + (D_{\rH}F_{\rL})^t\,\SD_v \PT{t}_{\rH} +
Q_{\rL}^t v_{\rL}^t \Bigr)\, dt 
\\
&\qquad +\Bigl((D_\rL Q_\rL)^t\,\SD_v\PT{t}_{\rL}\Bigr)\,dW_\rL(t)\, 
\\
&\qquad +\Bigl((D_\rH Q_\rL)^t\,\SD_v\PT{t}_{\rH}\Bigr)\,dW_\rL(t)\, 
,
\\[3mm]
d\,\SD_v \PT{t}_{\rH} &= \Bigl(-A_{\rH}\,\SD_v \PT{t}_{\rH} +
(D_{\rL}F_{\rH})^t \,\SD_v \PT{t}_{\rL} \label{e:DXH}\\
	& \qquad + (D_{\rH}F_{\rH})^t\,\SD_v \PT{t}_{\rH} +
Q_{\rH} v_\rH^t\Bigr)\,dt\;,
\end{equs}
with 
zero initial condition.
Since we want to consider derivatives with respect to the low-frequency part,
we would like to define (implicitly) $v_{\rH}^t$ as
\begin{equ}
v_{\rH}^t = -Q_{\rH}^{-1}(D_{\rL}F_{\rH})^t\,\SD_v
\PT{t}_{\rL}\;.
\end{equ}
In this way,  the solution of
\eref{e:DXH} would be $\SD_v \PT{t}_{\rH} \equiv 0$.
We next would define the ``directions'' $v_\rL$ and $v_\rH$ by
\begin{equa}
	v^t_\rL &= \bigl(V_\rL^t\,Q_\rL^t\bigr)^\transpose \;,\\
	v^t_\rH &= -Q_\rH^{-1} (D_\rL F_{\rH})^t 
\,\SD_v \PT{t}_{\rL}\;,
\end{equa}
where $\SD_v \PT{t}_{\rL}$ is the solution to \eref{e:DXL} with $\SD_v
\PT{t}_{\rH}$ replaced by $0$ and $v_\rL$ replaced by
$\bigl(V_\rL^t\,Q_\rL^t\bigr)^\transpose $. Here, $X^\transpose$
denotes the transpose of the real matrix $X$.
{\em This program needs some care because of the implicit
nature of the definition of $v$.}

Since we are constructing a solution of \eref{e:DX} whose
high-frequency part is going to vanish, we consider instead the
simpler equation for $y^t \in \SL(\CH_{\rL},\CH_\rL)$:
\begin{equ}[e:ext]
dy^t  = \Bigl(-A_{\rL} y^t + (D_{\rL} F_{\rL})^ty^t + Q_{\rL}^t \bigl(
V_\rL^tQ_\rL^t\bigr)^*\Bigr)\,dt+\bigl((D_\rL Q_\rL)^t y^t\bigr)\,dW_\rL(t)\;,
\end{equ}  
with $y^0=0$, and where we use again the notation $F^t=F\circ\Phi^t$,
and similar notation for 
$Q$.

The verification that \eref{e:ext} is well-defined and can be bounded
is again a consequence of \sref{theo:exist} and is left to the reader.
{\em Given the solution of \eref{e:ext} we proceed to make our
definitive choice of $v_\rL^t$ and $v_\rH^t$:} 
\begin{definition}
\label{def:vt}
Given an initial condition $ \ic \in\CH^\alpha $ (for $\Phi^t$)
and a cutoff $\rho <\infty $ we define $v^t=v_\rL^t\oplus v_\rH^t$ by
\begin{equa}[3][e:defu]
	v^t_\rL &\equiv\hphantom{-} \bigl(V_\rL^t\,Q_\rL^t\bigr)^\transpose
&&=\hphantom{-}\bigl(V_\rL^t\,(Q_\rL\circ\Phi^t)\bigr)^\transpose\;,\\
	v^t_\rH& \equiv -Q_\rH^{-1} (D_\rL F_\rH)^t\,y^t
&&= -Q_\rH^{-1} \bigl((D_\rL F_\rH)\circ\Phi^t)\,y^t\;,
\end{equa}
where  $\Phi^t$ solves
\eref{e:cutoff}, $V_\rL^t$ is the solution of \eref{e:Jac2}, and $y^t$
solves \eref{e:ext}. 
\end{definition}
\begin{lemma}
\label{lem:vh}The process $v^t$ satisfies for
all $p\ge 1$ and all $t>0$~:
\begin{equ}
\Exp \Bigl(\sup_{s \in [0,t]} \|v^s\|^p \Bigr)< C_{t,p,\rho}
\bigl(1+\nn{\alpha }{ \ic })^p\;,
\end{equ}
\ie, it satisfies assumption {\bf A2} of \sref{prop:parts}.
\end{lemma}

\begin{proof}By \sref{prop:reg}, $\Phi^t$ is in $\CH^\alpha $ for all
$t\ge0$. In \sref{lem:prop}~{\bf P6}, it will be checked that
$D_\rL F_\rH$ maps $\CH^\alpha $ into $\SL(\CH_\rL, \CH^\alpha \cap
\CH_\rH)$ and that this map has linear growth.
By the {\em lower bound} \eref{e:coeff} on the amplitudes $q_k$, we
see that $Q_\rH^{-1}$ is bounded from $\CH^\alpha \cap \CH_\rH$ to
$\CH_\rH$ and thus the assertion follows.
\end{proof}
We now verify that  $\SD_v \PT{t}_{\rH} \equiv 0$. Indeed, consider
the equations \eref{e:DX}. This is a system for two unknowns, $Y^t= \SD_v
\PT{t}_{\rL}$ and $X^t=\SD_v
\PT{t}_{\rH}$. For our choice of $v_\rL^t$ and $v_\rH^t$ this system
takes the form
\minilab{e:DXX}
\begin{equs}
d\,Y^t &= \Bigl(-A_{\rL}\,Y^t +
(D_{\rL}F_{\rL})^t \,Y^t \label{e:DXXL}
\\
	& \qquad + (D_{\rH}F_{\rL})^t\,X^t +
Q_{\rL}^t (V_\rL^tQ_\rL^t)^* \Bigr)\, dt 
\\
&\qquad +\Bigl((D_\rL Q_\rL)^t\,Y^t\Bigr)\,dW_\rL(t)\, 
\\
&\qquad +\Bigl((D_\rH Q_\rL)^t\,X^t\Bigr)\,dW_\rL(t)\, 
,
\\[3mm]
d\,X^t &= \Bigl(-A_{\rH}\,X^t +
(D_{\rL}F_{\rH})^t \,Y^t \label{e:DXXH}\\
	& \qquad + (D_{\rH}F_{\rH})^t\,X^t -(D_\rL F_\rH)^t y^t
\Bigr)\,dt\;.
\end{equs}
By inspection, we see that $X^t\equiv0$ 
and
\begin{equ}[e:varPhi]
dY^t = -A_{\rL}Y^t +
(D_{\rL}F_{\rL})^t Y^t +\bigl((D_\rL Q_\rL)^t\,Y^t\bigr)\,dW_\rL(t)
+ Q_{\rL}^t (V_\rL^tQ_\rL^t)^* \, dt 
\end{equ}
solve the problem, \ie, $Y^t=y^t$, by the construction of $y^t$.
Applying the It\^o formula to the product $V_\rL^t Y^t$ and using
Eqs.\eref{e:Jac2} and \eref{e:varPhi}, we see immediately  
that we have defined $Y^t=\SD_v
\PT{t}_{\rL}$ in such a way that
\begin{equ}
d \bigl(V_\rL^t \SD_v \PT{t}_{\rL}\bigr) = V_\rL^t Q_\rL^t
(Q_\rL^t)^*
(V_\rL^t)^\transpose \,dt\;,
\end{equ}
because all other terms cancel.
Thus we finally have shown
\begin{theorem}Given an initial condition $ \ic \in\CH^\alpha $ (for $\Phi^t$)
and a cutoff $\rho <\infty $, the following is true:
If $v^t$ is given by \sref{def:vt}
then
\begin{equa}[e:malliavin]
\SD_v \PT{t}_{\rL} &= U_\rL^t \int_0^t V_\rL^s \bigl((Q_\rL Q^\transpose_\rL)
\circ\Phi^s\bigr ) \bigl(V_\rL^s\bigr)^\transpose \,ds \equiv
U_\rL^t C_\rL^t\;,\\
\SD_v \PT{t}_{\rH}&\equiv 0\;.
\end{equa}
\end{theorem}

\begin{definition}
We will call the matrix $C_\rL^t$ the {\em partial Malliavin matrix} of our
system. 
\end{definition}

\section{The Partial Malliavin Matrix}

In this section, we estimate the partial Malliavin matrix $C_\rL^t$ from below.
We fix some time $t>0$ and denote by $\CS^L$ the unit sphere in
$\R^L$. Our bound is

\begin{theorem}
\label{theo:Cp}
There are constants $\mu,\nu \ge 0$ such that
for every $T>0$ and every $p\ge 1$ there is
a $C_{T,p,\rho}$
such that for all initial conditions $ \ic \in \SH^\alpha $ for the flow $\Phi^t$
and all $t< T$, one has
\begin{equ}
\Exp \Bigl(\bigl(\det C_\rL^t\bigr)^{-p} \Bigr )\le C_{T,p,\rho}
t^{-\mu p} \bigl(1 + \| \ic \|_\alpha  \bigr)^{\nu p}\;.
\end{equ}
\end{theorem}

\begin{corollary}
\label{cor:Cp}
There are constants $\mu,\nu \ge 0$ such that
for every $T>0$ and every $p\ge 1$ there is
a $C_{T,p,\rho}$
such that for all initial conditions $ \ic \in \SH^\alpha $ for the flow $\Phi^t$
and all $t< T$, one has, with $v$ given by \sref{def:vt}:
\begin{equ}
\Exp \nn[b]{}{\bigl(\SD_v \Phi^t_\rL\bigr)^{-p}}\le C_{T,p,\rho}
t^{-\mu p} \bigl(1 + \| \ic \|_\alpha  \bigr)^{\nu p}\;.
\end{equ}
\end{corollary}

This corollary follows from $(\SD_v \Phi^t_\rL)^{-1}=(C_\rL^t)^{-1}
V_\rL^t$ and Eq.\eref{e:VLa}. 

As a first step, we formulate a bound from which \sref{theo:Cp}
follows easily.
\begin{theorem}
\label{theo:covariance0}
There are a $\mu>0$ and a $\nu>0$ such that
for every $p\ge1$, every $t<T$ and every $ \ic  \in
\SH^2$, one has
\begin{equ}
\Prob\biggl(\inf_{h \in \CS^L} \int_0^t \bigl\|Q_\rL^s
\bigl(V_\rL^s\bigr)^\transpose  
h\bigr\|^2 \, ds < \eps\biggr) \,\le\, C_{T,p,\rho} \eps^p t^{-\mu p}
(1+\| \ic \|_2)^{\nu p}\;,
\end{equ}
with $C_{T,p,\rho}$ independent of $ \ic $.
\end{theorem}
\begin{proof}[of \theo{theo:Cp}]
Note that $\int_0^t \|Q_\rL^s
(V_\rL^s)^\transpose  
h\|^2\,ds$ is, by Eq.\eref{e:malliavin}, nothing but the quantity $\scal{h, C_\rL
^t h}$. Then, \sref{theo:Cp} follows at once.
\end{proof}

The proof of \sref{theo:covariance0}
is largely inspired from \cite[Sect.~4]{Norr}, but we
need some new features to deal with the infinite dimensional
high-frequency part. This will take up the next three subsections.

Our proof needs a modification of the Lie brackets considered when
we study the H\"ormander condition. We
explain first these identities in a finite dimensional setting.

\subsection{Finite Dimensional Case}

Assume in this subsection that both  $\CH_\rL$ and $\CH_\rH$ are finite dimensional.

The operator $Q$ maps $\CH$ to $\cscr L(\CW,\CH)$, where $\CW$ should
be thought of as the subspace of $\CH$ which is actually driven by the
noise.
We assume it is spanned by the first $W$ basis vectors of $\CH$.
We denote by $Q_i :\CH \to \CH$ the $i^{\rm th}$ column of $Q$,
($i=0,\dots,\dimI-1$).\footnote{There is a slight ambiguity of notation
here, since $Q_i$ really means $Q_{\rho,i}$ which is not the same as
$Q_\rho$.}
Finally, $\hF$ is the drift (in this section, we absorb the linear part
of the SDE into $\hF=-A+F$, to simplify the expressions).
The equation for $\PT{t}$ is
\begin{equ}
\PT{t}( \ic ) = \ic + \int_0^t \bigl(\hF  \circ\PT{s}\bigr)( \ic )\,ds + \int_0^t
\sum_{i=0}^{\dimI-1}\bigl(Q_i\circ\Phi^s\bigr)( \ic )\,dw_i(s)\;.
\end{equ}

Let $K : \CH \to \CH_\rL$ be a smooth function whose derivatives
are all
bounded and define  $K^t = K \circ \PT{t}$, $\hF^t=\hF\circ\Phi^t$,
and $Q_i^t=Q_i\circ
\Phi^t$.
We then have by It\^o's
formula
\begin{equ}[e:dKt]
d K^t = (D K )^t\hF^t\, dt +\sum_{i=0}^{\dimI-1}
(D K )^tQ_i^t \,dw_i(t) 
+\HALF\sum_{i=0}^{\dimI-1}(D^2 K)^t (Q_i^t;Q_i^t)\,dt\;.
\end{equ}
We next rewrite the equation \eref{e:Jac} for $V_\rL^t$ as:
\begin{equ}
dV_\rL^t = - V_\rL^t (D_\rL  \hF_\rL )^t \,dt- 
\sumiL V_\rL^t (D_\rL Q_i)^t \,dw_i(t)
+\sumiL V^t_\rL \bigl((D_\rL Q_i)^t\bigr)^2  
\,dt\;.
\end{equ}

The following definition is useful. Let $A : \CH
\to \CH$ and $B : \CH \to \CH_\rL$ be two functions with continuous bounded
derivatives. We define the {\em projected Lie bracket} $[A,B]_\rL :
\CH \to \CH_\rL$ by 
\begin{equ}{}
[A,B]_\rL(x) = \Pi_\rL [A,B](x) = \bigl( DB(x)\bigr) A(x) - \bigl( D_\rL A_\rL
(x)\bigr) B(x)\;.
\end{equ}

By It\^o's formula, we have therefore
the following equation for the product
$V_\rL^t K^t$:
\begin{equs}
d\bigl(V_\rL^t K^t\bigr) &= 
-V_\rL^t (D_\rL \hF _\rL)^t K^t\,dt
-V_\rL^t\sumiL (D_\rL Q_i)^t K^t\,dw_i(t)\\
&~~+V_\rL^t \sumiL \bigl((D_\rL Q_i)^t \bigr)^2 K^t\,dt 
 + V_\rL^t (D K)^t \hF^t  \, dt \\
&~~+ V_\rL^t\sum_{i=0}^{\dimI-1}(D K )^tQ_i^t\,dw_i(t) \\
&~~+\HALF V_\rL^t\sum_{i=0}^{\dimI-1}(D^2 K)^t  (Q_i^t;Q_i^t 
)\,dt \\
&~~- V_\rL^t\sumiL  (D_\rL Q_i)^t(D K)^t Q_i^t\,dt\;.
\end{equs}
By construction, $D_\rL Q_i=0$ for $i>L$ and therefore we extend all
the sums above to $W$. A straightforward calculation then leads to
\begin{equs}
d\bigl(V_\rL^t K^t\bigr) &= 
V_\rL^t \Bigl([\hF ,K]_\rL^t 
+ \HALF\sum_{i=0}^{\dimI-1}
\bigl[Q_i,[Q_i,K]_\rL\bigr]_\rL^t\Bigr)\,dt \label{e:dVK}\\
&~~+ V_\rL^t \sum_{i=0}^{\dimI-1}[Q_i,K]_\rL^t \,\,\,dw_i(t)\\
&~~ +\HALF V_\rL^t \sum_{i=0}^{\dimI-1}
\Bigl(\bigl((D_\rL Q_i )^t\bigr)^2 K^t
-(DK)^t(DQ_i)^tQ_i^t\\
&~~~~~~~~~~~~~~~~~~~~+( D D_\rL Q_i)^t(Q_i^t;K^t)
\Bigr)\;.
\end{equs}
Note next that for 
$i< L$, both $K$ and $Q_i$ map to $\CH_\rL$ and therefore $D D_\rL
Q_i(Q_i;K)=
D_\rL^2 Q_i(Q_i;K)$ when $i< L$ and it is $0$ otherwise.
Similarly, $(DK)(DQ_i)Q_i$ equals $(DK)(D_\rL Q_i)Q_i$ when $i< L$ and
vanishes otherwise. Thus, the last sum in \eref{e:dVK} only extends to $L-1$.

In order to simplify \eref{e:dVK} further, we define the vector field
$\tF  : \CH \to \CH$ by 
\begin{equ}
\tF  = \widehat F -\HALF\sumiL (D_\rL Q_i) Q_i\;.
\end{equ}
Then we get
\begin{equs}
d\bigl(V_\rL^t K^t\bigr) &= V_\rL^t \Bigl([\tF ,K]_\rL^t +
\HALF \sum_{i=0}^{\dimI-1}
\bigl[Q_i,[Q_i,K]_\rL\bigr]_\rL^t\Bigr)\,dt   
+ V_\rL^t\sum_{i=0}^{\dimI-1}  [Q_i,K]_\rL^t\,dw_i(t)\;.
\end{equs}
This is very similar to \cite[p.~128]{Norr}, who uses the conventional
Lie
bracket instead of $[\cdot,\cdot]_\rL\,$.

\subsection{Infinite Dimensional Case}

In this case, some additional care is needed 
when we transcribe \eref{e:dKt}. The
problem is that the stochastic flow $\PT{t}$ solves \eref{e:SDE} in the mild
sense but not in the strong sense. 
Nevertheless, this technical difficulty will
be circumvented by choosing the initial condition in $\CH^\alpha $. 
We have indeed by \sref{prop:reg}~\myitem{A} that
if the initial condition is in $\CH^\gamma$ with
$\gamma\in[1,\alpha]$, then the solution of \eref{e:SDE} is in the
same space. 
Thus, \sref{prop:reg} allows us to use It\^o's formula also in the
infinite dimensional case.

For any two Banach (or Hilbert) spaces $\cscr B_1$,
$\cscr B_2$, we denote by $\poly(\cscr B_1,\cscr B_2)$ the set of all
$\CC^\infty$ functions 
$\cscr B_1\to\cscr B_2$, which are polynomially bounded together with all their
derivatives. 
Let $K \in \poly(\CH,\CH_\rL)$ and $X \in \poly(\CH,\CH)$. We define
as above
$[X,K]_\rL \in
\poly(\CH,\CH_\rL)$ by
\begin{equ}{}
[X,K]_\rL(x) = \bigl( DK(x)\bigr) X(x) - \bigl( D_\rL X_\rL (x)\bigr) K(x)\;.
\end{equ}
Furthermore, we define $[A,K]_\rL \in \poly(\CD(A),\CH_\rL)$ by the
corresponding formula, \ie,
\begin{equ}{}
[A,K]_\rL(x) = \bigl( DK(x)\bigr)A x - A_\rL K(x)\;,
\end{equ}
where $A=1-\Delta$.
Notice that if $K$ is a constant vector field, \ie, $D K = 0$, then $[A,K]_\rL$
extends uniquely to an element of $\poly(\CH,\CH_\rL)$.

We choose again the basis $\{e_i\}_{i=0}^\infty$ of Fourier modes in
$\CH$ (see Eq.\eref{e:basis}) and define
$dw_i(t) = \scal{e_i,dW(t)}$. We also define the stochastic process
$K^t( \ic ) = (K \circ \Phi^t)( \ic )$ and 
\begin{equ}
\tF = F-\HALF \sumiL (D_\rL Q_i)Q_i\;,
\end{equ}
where $Q_i(x)= Q(x) e_i$.
Then one has
\begin{proposition}
\label{prop:evol}
Let $ \ic  \in \SH^1$ and $K\in \poly(\CH,\CH_\rL)$.  Then the equality
\begin{equs}
V_\rL^t( \ic ) K^t( \ic ) &= K( \ic ) + \int_0^t V_\rL^s ( \ic )\sum_{i=0}^\infty 
[Q_i,K]^s_\rL ( \ic )\,dw_i(s) \\
&\qquad+ \int_0^t V_\rL^s( \ic ) \Bigl(-[A,K]^s_\rL ( \ic ) +
[\tF,K]^s_\rL( \ic )\Bigr)\,ds\\
&\qquad+ \HALF \int_0^t
V_\rL^s( \ic )\sum_{i=0}^\infty\bigl[Q_i,[Q_i,K]_\rL\bigr]_\rL^s ( \ic )\,ds\;,
\end{equs}
holds almost surely. Furthermore, the
same equality holds if $ \ic  \in \SH^2$ and $K\in \poly(\SH^1,\CH_\rL)$.
\end{proposition}
Note that by $[A,K]_\rL^s( \ic )$ we mean
$\Bigl(DK\bigl(\Phi^s( \ic )\bigr)\Bigr)
\bigl( A
\Phi^s( \ic )\bigr)- 
A_\rL K\bigl(\Phi^s( \ic )\bigr)$.
\begin{proof}
It\^o's formula.
\end{proof}

\subsection{The Restricted H\"ormander Condition}

The condition for having appropriate mixing properties is the following
H\"orman\-der-like condition.
\begin{definition}
Let $\CK = \{K^{(i)}\}_{i=1}^M$ be a collection
of functions in $\poly(\CH,\CH_\rL)$.
We say that $\CK$ satisfies the \emph{restricted H\"ormander
condition} if
there exist constants $\delta, R>0$ such that for every
$h\in
\CH_\rL$ and every $y \in \CH$ one has
\begin{equ}[e:Horm]
\sup_{K \in \CK}\,\, \inflikesup_{\|x - y\| \le R} \scal{h,K(x)}^2 \ge \delta \|h\|^2\;.
\end{equ}
\end{definition}

We now construct the set $\CK$ for our problem.
We define the operator $[X^0,\cdot\;]_\rL : \poly(\CH^\gamma ,\CH_\rL) \to
\poly(\CH^{\gamma +1},\CH_\rL)$ by
\begin{equ}{}
[X^0,K]_\rL = -[A, K]_\rL + [F, K]_\rL + \HALF \sum_{i=0}^\infty
\bigl[Q_i,[Q_i, K]_\rL\bigr]_\rL-\HALF\sumiL \bigl[(D_\rL Q_i)
Q_i,K\bigr]_\rL \;.
\end{equ}
This is a well-defined operation since  $Q$ is Hilbert-Schmidt and
$DQ$ is finite rank and
we can write
\begin{equ}
\sum_{i=0}^\infty \bigl[Q_i,[Q_i, K]_\rL\bigr]_\rL = \sum_{i=0}^\infty
\bigl(D^2K\bigr)(Q_i;Q_i) + r\;,
\end{equ}
with $r$ a finite sum of bounded terms.

\begin{definition}
\label{def:K}
We define
\begin{claim}
\item[--]$\CK_0=\{ Q_i,\hbox{~with~} i=0,\dots,L-1 \}$,
\item[--]$\CK_1=\{ [X^0,Q_i]_\rL, \hbox{~with~} i=\kzero ,\dots,L-1 \}$,
\item[--]$\CK_\ell=\{ [Q_i,K]_\rL, \hbox{~with~}
K\in\CK_{\ell-1}\hbox{~and~}i=\kzero ,\dots,L-1 \}$, when $\ell>1$.
\end{claim}
Finally,
\begin{equ}
\CK=\CK_0 \cup \cdots\cup \CK_3\;.\footnote{The number 3 is the power
3 in $u^3$.}
\end{equ}
\end{definition}
\begin{remark}
\label{rem:notsobad}
Since for $i\ge \kzero $ the $Q_i$ are constant vector fields,
the quantity $[X^0,K]$ is in $\poly(\CH,\CH_\rL)$ and not only in
$\poly(\CH^1,\CH_\rL)$.
Furthermore, if $K\in\CK$ then $D^j K$ is bounded for all $j\ge0$.
\end{remark}

We have
\begin{theorem}
\label{theo:restricted} The set $\CK$ constructed above
satisfies the
restricted H\"ormander condition for the cutoff\, GL equation
if $\rho$ is chosen sufficiently large. Furthermore, the inequality
\eref{e:Horm} holds for $R=\rho/2$. Finally, $\delta >\delta _0>0$ for
all sufficiently large $\rho $. 
\end{theorem}

\begin{proof}
The basic idea of the proof is as follows: The leading term of $F$ is
the cubic term $u^m$ with $m=3$. Clearly, if $i_1$, $i_2$, $i_3$ are
any 3 modes, we find
\begin{equ}[e:pm]
\bigl[e_{i_1},[e_{i_2},[u\mapsto u^3,e_{i_3}]_\rL]_\rL\bigr]_\rL =
 \sum_{k=\pm i_1\pm i_2 \pm i_3} C_k \Pi_\rL e_k\;,
\end{equ}
where the $e_\ell$ are the basis vectors of $\CH$ defined in 
\eref{e:basis}, 
and the $C_k$ are
{\em non-zero} combinatorial constants. 
By our \sref{def:i} of the set $\I$, the following is true: For every
choice of a fixed $k$ the three numbers $i_1$, $i_2$, and $i_3$ of
$\I_k$
satisfy
\begin{claim}
\item[--] For $j=1,2,3$ one has $ i_j\in \{ \kzero ,\dots,L-1\}$.
\item[--] If $|k|< \kzero $ exactly one of the six sums 
$\pm i_1\pm i_2 \pm i_3$ lies in the set
$\{0,\dots,\kzero-1 \}$ and exactly one lies in $\{-(\kzero-1) ,\dots,0\}$.
\end{claim}
In particular, the
expression \eref{e:pm} does not depend on $u$. If instead of $u^3$ we
take a lower 
power, the triple commutator will vanish. 

The basic idea has to be slightly modified because of the cutoff
$\rho$. First of all, the constant $R$ in the definition of the
H\"ormander condition is set to $R=\rho/2$.
Consider first the case where $\nn{}{x}\ge 5\rho/2 $. In that case we
see from \eref{e:Qrho} that the $Q_{\rho,i}$, viewed as vector fields,
are of
the form
\begin{equ}
Q_{\rho,i}(x)\,=\,\left \{
\begin{array}{rl}
(q_i+1)e_i,& \text{if $i<\kzero $,}\\
q_ie_i,& \text{if $i\ge \kzero $.}
\end{array}
\right .
\end{equ}
Since these vectors span a basis of $\CH_\rL$ the inequality
\eref{e:Horm} follows in this case (already by choosing only $K\in\CK_0$).

Consider next the more delicate case when $\nn{}{x}\le 5\rho/2 $.
\begin{lemma}
\label{lem:h0}For all $\nn{}{x}\le 3\rho  $ one has
for $\{i_1,i_2,i_3\}=\I_k$ the identity
\begin{equ}[e:pm2]
\bigl[e_{i_1},[e_{i_2},[X^0,e_{i_3}]_\rL]_\rL\bigr]_\rL(x) =
 \sum_{k=\pm i_1\pm i_2 \pm i_3} C_k \Pi_\rL e_k + r_\rho(x)\;,
\end{equ}
where $r_\rho$ satisfies a bound
\begin{equ}
\nn{}{r_\rho(x)}\,\le\,
C \rho ^{-1}\;,
\end{equ}
with the constant $C$ independent of $x$ and of $k< \kzero $.
\end{lemma}

\begin{proof}
In $[X^0,\cdot]_\rL$ there are 4 terms. 
The first, $A$, leads successively to
$[A,e_{i_3}]_\rL= (1+i_3^2)e_{i_3}$, which is constant, and hence the
Lie bracket with $e_{i_2}$ vanishes.
The second term contains the non-linear interaction $\FR$.
Since $\nn{}{x}\le 3\rho  $ 
one has $\FR(x)=F(x)$.
Thus, \eref{e:pm} yields the leading term of \eref{e:pm2}.
The two remaining terms will contribute to $r_\rho(x)$. We just
discuss the first one.
We have, using \eref{e:Qrho}, 
\begin{equ}
{}[Q_{\rho,i},e_{i_3}]_\rL(x)= -DQ_{\rho,i}(x)
e_{i_3}\,=\,-{1\over \rho}\chi'(\nn{}{x}/\rho){\scal{x,e_{i_3}} \over
\nn{}{x}}
\Pi_{\kzero }e_i\;.
\end{equ}
This gives clearly a bound of order $\rho ^{-1}$ for this Lie bracket,
and the further ones are handled in the same way.
\end{proof}

We continue the proof of \sref{theo:restricted}. 
When $k<k_*$, we consider the elements of $\CK_3$. They are of the
form
\begin{equ}
\bigl[Q_{\rho,i_1},[Q_{\rho ,i_2},[X^0,Q_{\rho ,i_3}]_\rL]_\rL\bigr]_\rL(x) =
q_{i_1}q_{i_2}q_{i_3}\Bigl( \sum_{k=\pm i_1\pm i_2 \pm i_3} C_k \Pi_\rL e_k + r_\rho(x)\Bigr)\;.
\end{equ}
Thus, for $\rho =\infty $ these vectors together with the $Q_i$ with
$i\in \{k_*,\dots,L-1\}$ span $\CH_\rL$ (independently of $y$ with
$\nn{}{y}\le 3\rho $)
and therefore \eref{e:pm} holds in this case, if $\|x\|\le 5\rho /2$
and $R=\rho/2$. The assertion for finite, but large enough $\rho $
follows immediately by a perturbation argument.
This completes the case of  $\|x\|\le 5\rho /2$ and hence the proof of
\sref{theo:restricted}. 
\end{proof}

\begin{proof}[of \theo{theo:covariance0}] 
The proof is very similar to the one in \cite{Norr}, but we have to
keep track of the $x,t$-dependence of the estimates. 
First of all, choose $x\in \CH^2$ and $t \in (0, t_0]$.

{}From now on, we will use the notation $\CO(\nu)$ as a shortcut for
$C(1+\|x\|_2^{\nu})$, where the constant $C$ may
depend on $t_0$ and $p$, but is independent of $x$ and $t$. 
Denote by $R$ the constant found in \sref{theo:restricted} and
define the subset  $\CB_x$ of $\CH^2$ by 
\begin{equ}
\CB_x = \bigl\{y \in \CH^2 \;:\; \|y - x\| \le R~~~\text{and}~~~\|y\|_\gamma \le \|x\|_\gamma + 1 \;\text{for $\gamma = 1,2$}\bigr\}\;.
\end{equ}
We also denote by $\CB(I)$ a ball of (small) radius $\CO(1/L)$
centered at the identity in the space of all $L\times L$
matrices. (Recall that $L$ is the dimension of $\CH_\rL$, and that $K \in \CK$
maps to $\CH_\rL$.)
We then have
a bound of the type
\begin{equ}[e:cond1]
\sup_{y \in \CB_x}\;\sup_{K \in \CK}\sum_{i=0}^\infty \bigl\|[Q_i, K]_\rL
(y)\bigr\|^2 \le \CO(0)\;.
\end{equ}

This is a consequence of the fact that $QQ^\transpose$ is trace
class and thus the sum converges and its principal term is equal to
\begin{equa}
\tr \bigl(Q^\transpose(y)\,& \bigl(DK\bigr)^\transpose(y)\,
\bigl(DK\bigr)(y)\, Q(y)\bigr) \\&= 
\tr \bigl(\bigl(DK\bigr)(y) Q(y)\,Q^\transpose(y)\,
\bigl(DK\bigr)^\transpose(y) 
\bigr) \\ 
&= \sumiL  \|Q^\transpose(y)\,\bigl(DK\bigr)^\transpose(y)e_i\|^2 \le C_\rho\;.
\end{equa}
The last inequality follows from \sref{rem:notsobad}. The other terms
form a finite sum containing derivatives of the $Q_i$ and are bounded in
a similar way.

We have furthermore bounds of the type
\begin{equa}[e:cond2]
\sup_{y \in \CB_x}\;\sup_{K \in \CK}\; \bigl\|[X^0,K]_\rL(y) \bigr\|^2 &\le
\CO(\nu)\;,\\
\sup_{y \in \CB_x}\;\sup_{K \in \CK}\;
\bigl\|\bigl[X^0,[X^0,K]_\rL\bigr]_\rL(y) \bigr\|^2 &\le \CO(\nu)\;,\\
\sup_{y \in \CB_x}\;\sup_{K \in \CK}\; \sum_{i=0}^\infty
\bigl\|\bigl[Q_i,[X^0,K]_\rL\bigr]_\rL(y) \bigr\|^2 &\le \CO(\nu)\;,
\end{equa}
where $\nu=1$.

Let $\CS_\rL$ be the unit sphere in $\CH_\rL$. By the assumptions on
$\CK$ and the choice of $\CB(I)$ we see that:
\begin{claim}
\item[\myitem{A}]\it For every $\vzero \in \CS_\rL$, there exist a $K \in \CK$ and a
neighborhood $\CN$ of $\vzero $ in $\CS_\rL$ such that
\begin{equ}
\inf_{y \in \CB_x}\inf_{V \in \CB(I)}\inf_{\uzero  \in \CN} \scal{\,V K(y),\uzero }^2 \ge
{\delta\over 2}\;,
\end{equ}
with $\delta$ the constant appearing in \eref{e:Horm}.
\end{claim}
Next, we define a stopping time $\tau $ by $\tau =
\min\{t,\tau_1,\tau_2\}$ with 
\begin{equs}
\tau_1 &= \inf\bigl\{s \ge 0\;:\; \Phi^s(x)\not\in \CB_x \bigr\}\;,\\
\tau_2 &= \inf\bigl\{s \ge 0\;:\; V_\rL^s(x) \not\in \CB(I) \bigr\}\;,\\
t&<T \text{ as chosen in the statement of \theo{theo:covariance0}}\;.
\end{equs}
It follows easily from \sref{prop:reg}~\myitem{E} that the probability of $\tau_1$ being small (meaning that
in the sequel we will always assume $\eps \le 1$) can be bounded by
\begin{equ}
\Prob(\tau_1 < \eps) \le C_p (1+\|x\|_2)^{16 p} \eps^p\;,
\end{equ}
with $C_p$ independent of $x$.
Similarly, using \sref{lem:VL}, we see that
\begin{equ}
\Prob(\tau_2 < \eps) \le C_p \eps^p\;.
\end{equ}
Observing that $\Prob(t<\epsilon )<t^{-p}\epsilon ^p$ and combining this
with the two estimates, we get for every $p\ge1$:
\begin{equ}
\Prob\bigl(\tau  < \eps\bigr) \le \CO(16p)t^{-p} \eps^p\;.
\end{equ}
From this and \myitem{A}~we deduce
\begin{claim}
\item[\myitem{B}]\it for every $\vzero \in \CS_\rL$ there exist a $K \in \CK$ and a
neighborhood $\CN$ of $\vzero $ in $\CS_\rL$ such that for $\eps < 1$,
\begin{equ}
\sup_{\uzero  \in \CN} \Prob \biggl( \int_0^\tau \scal[b]{V_\rL^s(x)
K^s(x),\uzero }^2 \,ds\le \eps\biggr) \le \Prob(\tau <
2\eps/\delta) 
\le
\CO(16p)t^{-p} \eps^p\;,
\end{equ}
with $\delta$ the constant appearing in \eref{e:Horm}.
\end{claim}
Following \cite{Norr}, we will show below that \myitem{B}~implies:
\begin{claim}
\item[\myitem{C}]\it for every $\vzero \in \CS_\rL$ there exist an 
$i \in \{k_*,\ldots,L-1\}$, a
neighborhood $\CN$ of $\vzero $ in $\CS_\rL$ and constants $\nu,\mu>0$ such that for
$\eps<1$ and $p>1$ one has
\begin{equ}
\sup_{\uzero  \in \CN} \Prob \biggl( \int_0^\tau \scal[b]{V_\rL^s(x) Q_i^s(x),\uzero }^2 \,ds\le
\eps\biggr)  \le \CO(\nu p)t^{-\mu p} \eps^p\;.
\end{equ}
\end{claim}
\begin{remark}Note that for small $\nn{}{x}$,
$Q_i(x)=Q_{i,\rho}(x)$ may be 0 when $i<k_*$, but the point is that
then we can find another $i$ for which the inequality holds.
\end{remark} 
By a straightforward argument, given in detail in \cite[p.~127]{Norr}, 
one concludes that \myitem{C} implies
\theo{theo:covariance0}.

It thus only remains to show that \myitem{B}~implies \myitem{C}.
We follow closely Norris and choose a $K\in \CK$ such that
\myitem{B}~holds. If $K$ happens to be in $\CK_0$ then it is equal to
a $Q_i$, and thus we already have \myitem{C}. Otherwise, assume
$K\in\CK_j$ with $j\ge1$. Then we use a Martingale inequality. 
\begin{lemma}
\label{lem:mart}
Let $\CH$ be a separable Hilbert space and $W(t)$ be the cylindrical Wiener
process on $\CH$. Let $\beta^t$ be a real-valued predictable process and
$\gamma^t$, $\zeta^t$ be predictable $\CH$-valued processes. Define
\begin{equa}
a^t &= a^0 + \int_0^t \beta^s\,ds + \int_0^t \scal{\gamma^s,dW(s)}\;,\\
b^t &= b^0 + \int_0^t a^s\,ds + \int_0^t \scal{\zeta^s,dW(s)}\;.
\end{equa}
Suppose $\tau \le t_0$ is a bounded stopping time such that for some
constant $C_0 
< \infty$ we have
\begin{equ}
\sup_{0<s<\tau} \bigl\{|\beta^s|,\,|a^s|,\,\|\zeta^s\|,\,\|\gamma^s\|\bigr\} \le
C_0\;.
\end{equ}
Then, for every $p>1$, there exists a constant $C_{p,t_0}$ such that
\begin{equ}
\Prob\Bigl(\int_0^\tau (b^s)^2\,ds < \eps^{20} \quad\text{\rm and}\quad
\int_0^\tau \bigl(|a^s|^2 + \|\zeta^s\|^2\bigr)\,ds \ge \eps\Bigr) \le
C_{p,t_0}\,(1+C_0^6)^p\eps^p\;,
\end{equ}
for every $\eps \le 1$.
\end{lemma}
\begin{proof}
The proof is given in \cite{Norr}, but without the explicit dependence on
$C_0$. If we follow his proof carefully we get an estimate of the type
\begin{equ}
\Prob\Bigl(\int_0^\tau (b^s)^2\,ds < \eps^{10} \;\text{\rm and}\;
\int_0^\tau \bigl(|a^s|^2 + \|\zeta^s\|^2\bigr)\,ds \ge (1+C_0^3)\eps\Bigr) \le
C_1\,(1+C_0^{12})^p\eps^p\;.
\end{equ}
Replacing $\eps$ by $\eps^2$ and making the assumption $\eps < 1/(1+C_0^3)$, we
recover our statement. The statement is trivial for $\eps > 1/(1+C_0^3)$, since
any probability is always smaller than $1$.
\end{proof}
We apply this inequality as follows:
Define, for $K_0\in \CK$,
\begin{equs}
a^t(x) &= \scal[b]{V_\rL^t \bigl([X^0,K_0]_\rL^t\bigr)(x),\uzero }\;, \\
b^t(x) &= \scal[b]{V_\rL^t K_0^t(x),\uzero }\;,\\
\beta^t(x) &= \scal[b]{V_\rL^t \bigl(\bigl[X^0,[X^0,K_0]_\rL\bigr]_\rL^t\bigr)(x),\uzero }\;,\\
(\gamma^t)^i(x) &= \scal[b]{V_\rL^t \bigl(\bigl[Q_i,[X^0,K_0]_\rL\bigr]_\rL^t\bigr)(x),\uzero }\;,\\
(\zeta^t)^i (x) &= \scal[b]{V_\rL^t \bigl([Q_i,K_0]_\rL^t\bigr)(x),\uzero }\;.
\end{equs}
In this expression, $\zeta^t(x) \in \CH$, $(\zeta^t)^i(x) = \scal{\zeta^t(x),e_i}$
and similarly for the $\gamma^t$. 
It is clear by \sref{prop:evol}, Eq.\eref{e:cond1}, and
Eq.\eref{e:cond2} that the 
assumptions of \lem{lem:mart} are satisfied with $C_0 = \CO(\nu
)$ for some 
$\nu>0$. 

We continue the proof that \myitem{B} implies \myitem{C} in the case
when $K\in \CK_j$, with $j=1$.
Then, by the construction of $\CK_j$ with $j\ge1$,
there is a $K_0 \in \CK_{j-1}$ such that we have
either $K = [Q_i,K_0]_\rL$ for some $i \in \{\kzero ,\dots,L-1\}$, or $K =
[X^0,K_0]_\rL$. In fact, for $j=1$ only the second case occurs and
$K_0=Q_i$ for some $i$, but we are already preparing an inductive step.
Applying \lem{lem:mart}, we have for every $\epsilon
\le 1$:
\begin{equs}
\Prob\biggl( &\int_0^\tau \scal[b]{V_\rL^s K_0^s (x),\uzero }^2 \,ds < \eps
\quad\text{and}\quad \int_0^\tau \Bigl( \scal[b]{V_\rL^s [X^0,K_0]_\rL^s (x),\uzero }^2 \\
&+\sum_{i=0}^\infty  \scal[b]{V_\rL^s [Q_i,K_0]_\rL^s(x),\uzero }^2\Bigr)\,ds \ge \eps^{1/20} \biggr) \le \CO(6\nu
p)\eps^{p/20}\;.
\end{equs}
Since the second integral above
is always larger than
$
\int_0^\tau \scal[b]{V_\rL^t K^t (x),\uzero }^2 \,dt
$,
the probability for it to be smaller than $\eps^{1/20}$ is, by \myitem{B},
bounded by
$
\CO(16 p)t^{-p} \eps^{p/20}$.
This implies (replacing $\nu$ by $\max\{6\nu,16\}$) that
\begin{equ}
\Prob\Bigl(\int_0^\tau \scal[b]{V_\rL^s K_0^s (x),\uzero }^2 \,ds < \eps\Bigr) \le
\CO(\nu p)t^{-p}\eps^{p/20}\;.
\end{equ}
Since for $j=1$ we have $K_0=Q_i$ with $i\in\{k_*,\dots,L-1\}$, we
have shown \myitem{C} in this case. The above reasoning is repeated for
$j=2$ and $j=3$, by iterating the above argument. For example, if
$K=[Q_{i_1},[X^0,Q_{i_2}]_\rL]_\rL$ ,with
$i_1,i_2\in\{k_*,\dots,L-1\}$, we apply \lem{lem:mart} twice, showing
the 
first time 
that
$\scal{[X^0,Q_{i_2}]_\rL,\uzero }^2$ is unlikely to be small and then
again to show that $\scal{Q_{i_2},\uzero}^2$ is also unlikely to be
small (with other powers of $\epsilon $), which is what we wanted.
Finally, since every $K$ used in \myitem{B} is in $\CK$,
at most 3 such invocations of \lem{lem:mart} will be sufficient to
conclude that \myitem{C} holds.
The proof of \theo{theo:covariance0} is complete.
\end{proof}

\subsection{Estimates on the Low-Frequency Derivatives (Proof of 
\prop{prop:lowone})}

Having proven the crucial bound \sref{theo:Cp} on the reduced
Malliavin matrix,
we can now proceed to prove \sref{prop:lowone}, \ie,
the smoothing properties of the dynamics
in the low-frequency part.
For convenience, we restate it here.
\begin{proposition}
\label{prop:lowtwo}
There exist a time $T^*>0$ and exponents $\mu,\nu>0$ such that for every
$\phi \in \CC^2_b(\CH)$, every $ \ic \in\SH^\alpha$, and every $t\le T^*$,
one has
\begin{equ}[e:estDLphiL]
\Bigl\|\Exp \Bigl(\bigl(D_{\rL}\phi\circ\Phi^t\bigr)( \ic )(D_\rL
\Phi_\rL^t)(\ic)\Bigr)\Bigr\| \le C t^{-\mu} \bigl(1+\| \ic \|_\alpha^\nu\bigr)
\|\phi\|_{\L^\infty} \;.
\end{equ}
\end{proposition}
\begin{proof}
The proof will use the integration by parts formula 
\eref{e:parts}
together with \sref{theo:Cp}.
Fix $ \ic  \in \SH^\alpha$ and $t>0$. In this proof, we omit the argument
$ \ic $ 
to gain legibility, but it will
be understood that the formulas do generally only hold if evaluated at some
$ \ic \in \SH^\alpha$. We extend our phase space to include $D_\rL \Phi^t$, $V_\rL^t$ and
 $\SD_v \Phi^t_\rL$. We define a new stochastic process $\Psi^t$ by
\begin{equ}
\Psi^t = \bigl(\Phi^t, \SD_v \Phi^t_\rL, D_\rL \Phi^t, V_\rL^t\bigr) \in\WSH = \CH  \oplus \R^{L\cdot L} \oplus \CH^L\oplus \R^{L\cdot L} \;.
\end{equ}
Applying the definitions of these processes, we see that $\Psi^t$ is defined by the
autonomous SDE given by
\begin{equs}
d\Phi^t &= -A \Phi^t\,dt + F(\Phi^t)\, dt + Q(\Phi^t)\,dW(t)\;,\\
dD_\rL \Phi^t &= -A D_\rL \Phi^t\, dt + DF(\Phi^t) D_\rL \Phi^t\, dt +  DQ(\Phi^t) D_\rL\Phi^t\, dW(t)\;,\\
dV_\rL^t &= V_\rL^t A_\rL\,dt - V_\rL^t\, D_\rL F_\rL(\Phi^t)\,dt - V_\rL^t D_\rL Q_\rL(\Phi^t)\, dW_\rL(t) \\
&\qquad + V_\rL^t \sumiL  \bigl( D_\rL Q_\rL^i(\Phi^t)\bigr)^2\, dt\;,\\
d \SD_v \Phi_\rL^t &= -A_\rL \SD_v \Phi_\rL^t\, dt + D_\rL F_\rL(\Phi^t) \SD_v \Phi_\rL^t\,dt + Q_\rL(\Phi^t)^2(V_\rL^t)^*\,dt \\
&\qquad + D_\rL Q_\rL(\Phi^t)\,\SD_v \Phi_\rL^t\, dW_\rL(t)\;.
\end{equs}
This expression will be written in the short form
\begin{equ}
d\Psi^t = - \hA \Psi^t\, dt + \tF(\Psi^t)\, dt + \hQ(\Psi^t)\,dW(t)\;,
\end{equ}
with $\Psi^t \in \WSH$ and $dW(t)$ the cylindrical Wiener process on $\CH$.
It can easily be verified that this equation satisfies assumption {\bf A1} of \sref{prop:parts}. We consider again the stochastic process $v^t \in \CH$ defined in \eref{e:defu}.
It is clear from \sref{lem:vh} that $v^t$ satisfies {\bf A2}. With this particular choice of $v$, the first component of $\SD_v \Psi^t$ (the one in $\CH$) is equal to $\SD_v\Phi^t_\rL \oplus 0$.

We choose a function $\phi \in \CC_b^2(\CH)$ and fix two indices
$i,k \in \{0,\ldots,L-1\}$. Define $\tilde \phi_{i,k} : \WSH \to \R$ by 
\begin{equ}
\tilde\phi_{i,k}(\Psi^t) = \sum_{j=0}^{L-1} \phi(\Phi^t) 
\bigl((\SD_v \Phi^t_\rL)^{-1}\bigr)_{i,j}\bigl(D_\rL \Phi^t_\rL\bigr)_{j,k}\;,
\end{equ}
where the inverse has to be understood as the inverse of a square matrix.
By \sref{theo:Cp}, $\tilde\phi_{i,k}$ satisfies the assumptions of \sref{prop:parts}.
A simple computation gives for every $h \in \R^L$ the identity:
\begin{equs}
D\tilde \phi_{i,k}(\Psi^t) \SD_v \Psi^t h
&= D_\rL \phi (\Phi^t)\bigl(\SD_v \Phi_\rL^t h\bigr) \bigl((\SD_v
\Phi^t_\rL)^{-1}\bigr)_{i,j}\bigl(D_\rL \Phi^t_\rL\bigr)_{j,k} \\
&\quad + \phi(\Phi^t)\bigl( (\SD_v \Phi^t_\rL)^{-1}(\SD^2_v \Phi^t_\rL
h)(\SD_v \Phi^t_\rL)^{-1} \bigr)_{i,j}\bigl(D_\rL \Phi^t_\rL\bigr)_{j,k}\\
&\quad + \phi(\Phi^t)\bigl((\SD_v D_\rL\Phi_\rL^t)h\bigr)_{i,j}\bigl((\SD_v \Phi^t_\rL)^{-1}\bigr)_{j,k}\;,\label{e:xyz}
\end{equs}
where summation over $j$ is implicit.
We now apply the integration by parts formula in the form of \sref{prop:parts}.
This gives the identity
\begin{equs}
\Exp\bigl(D\tilde\phi_{i,k}(\Psi^t)\SD_v \Psi^t h \bigr)
= \Exp\Bigl(\tilde\phi_{i,k}(\Psi^t)\int_0^t
\scal[b]{v^s h,dW(s)}\Bigr)\;.
\end{equs}
Substituting \eref{e:xyz}, we find 
\begin{equs}
\Exp \Bigl(D_\rL \phi (\Phi^t)&\bigl(\SD_v \Phi_\rL^t h\bigr)  \bigl((\SD_v
\Phi^t_\rL)^{-1}\bigr)_{i,j}\bigl(D_\rL \Phi^t_\rL\bigr)_{j,k}\Bigr) = 
\\
&- \Exp \Bigl(\phi(\Phi^t)\bigl( (\SD_v \Phi^t_\rL)^{-1}(\SD^2_v \Phi^t_\rL
h)(\SD_v \Phi^t_\rL)^{-1} \bigr)_{i,j}\bigl(D_\rL \Phi^t_\rL\bigr)_{j,k}
\Bigr)
\\
&-\Exp\Bigl(\phi(\Phi^t)\bigl((\SD_v
D_\rL\Phi_\rL^t)h\bigr)_{i,j}\bigl((\SD_v \Phi^t_\rL)^{-1}\bigr)_{j,k}\Bigr)
\\
&+ \Exp\Bigl(\phi(\Phi^t) \bigl((\SD_v \Phi^t_\rL)^{-1}\bigr)_{i,j}\bigl(D_\rL \Phi^t_\rL\bigr)_{j,k}\int_0^t\scal[b]{v^s h,dW(s)}\Bigr) \;.
\end{equs}
The summation over the index $j$ is implicit in every term.
We now choose $h = e_i$ and sum over the index $i$. The left-hand side 
is then equal to 
\begin{equ}
\Exp \Bigl(\bigl(D_\rL \phi (\Phi^t)\bigr)D_\rL \Phi^t_\rL e_k\Bigr)\;,
\end{equ}
which is precisely the expression we want to bound. The right-hand
side can be bounded in terms of $\|\phi\|_{\L^\infty}$ and of $\Exp
\bigl((\SD_v \Phi^t_\rL)^{-4}\bigr)$ (at worst). The other factors are
all given by components of $\SD_v \Psi^t$ and can therefore be bounded
by means of \sref{theo:exist}. Therefore, 
\eref{e:estDLphiL} follows. The proof of
\sref{prop:lowtwo} is complete. 
\end{proof}
%
\section{Existence Theorems}
%
%
In this section, we prove existence theorems for several PDE's and
SDE's, in particular we prove \sref{prop:reg} and \sref{lem:cross}.
Much of the material here relies on well-known techniques, but
we include the details for completeness.

We consider again 
the problem
\begin{equ}[e:new1]
d\Phi^t =-A \Phi^t \,dt + F(\Phi^t) \,dt +
Q(\Phi^t)\, dW(t)\;,
\end{equ}
with $\Phi^0= \ic $ given.
The initial condition $ \ic $ will be taken in one of the
Hilbert spaces $\CH^\gamma $. We
will show that, after some time, the solution lies in some smaller Hilbert
space. Note that we are working here with the {\em cutoff} equations,
but we omit the index $\rho$.

We will of course require that all stochastic processes are
predictable. This means that if we write $\L^p(\Omega,\SY)$, with $\SY$ some
Banach space of functions of the interval $[0,T]$, we really mean that the
only functions we consider are those that are measurable with respect to the
predictable $\sigma$-field when considered as functions over $\Omega \times
[0,T]$.

We first state precisely
what is known about the ingredients of \eref{e:new1}. 
\begin{lemma}
\label{lem:prop}
The following properties hold for $A$, $F$ and $Q$.
\begin{claim}
\item[\bf P1]
The space $\CH$ is a real separable Hilbert space and $A: \CD(A) \to \CH$
is a self-adjoint strictly positive operator. 
\item[\bf P2] The map $F : \CH \to \CH$ has bounded derivatives of all orders.
\item[\bf P3] For every $\gamma \ge 0$, $F$ maps $\CH^\gamma$ into itself.
Furthermore, there exists a constant $n>0$ independent of $\gamma$ and
constants $C_{F,\gamma}$ such that
$F$ satisfies the bounds
\minilab{e:boundF}
\begin{equs}
\nn{\gamma}{F(x)}  &\le C_{F,\gamma}\bigl(1 +
\nn{\gamma}{x}\bigr)\;,\label{e:boundF1} \\
\nn{\gamma}{F(x) - F(y)}  &\le C_{F,\gamma}\nn{\gamma}{x-y}\bigl(1 +
\nn{\gamma}{x} + \nn{\gamma}{y}\bigr)^n\;,\label{e:boundF2}
\end{equs}
for all $x$ and $y$ in $\CH^\gamma$.

\item[\bf P4] There exists an $\alpha > 0$ such that for every $x,x_1,x_2 \in
\CH$ the map $Q:\CH \to \SL(\CH,\CH)$ satisfies
\begin{equ}
\bigl\|A^{\alpha-3/8} Q(x)\bigr\|_\HS \le C\;,\quad \bigl\|A^{\alpha-3/8}
\bigl(Q(x_1) - Q(x_2)\bigr)\bigr\|_\HS \le C\nn{}{x_1-x_2} \;,
\end{equ}
where $\|\cdot\|_\HS$ denotes the Hilbert-Schmidt norm in $\CH$.

\item[\bf P5] The derivative of $Q$ satisfies
\begin{equ}[e:P5]
\nn[b]{\HS}{A^\alpha \bigl(DQ(x)\bigr)h} \le C\nn{}{h}\;,
\end{equ}
for every $x,h \in \CH$.

\item[\bf P6] The derivative of $F$ satisfies 
\begin{equ}
\nn[b]{\gamma}{\bigl(DF(x)\bigr)y} \le C (1+\|x\|_\gamma) \|y\|_\gamma\;,
\end{equ}
for every $x,y \in \CH^\gamma$.
\end{claim}
\end{lemma}
\begin{proof}
The points {\bf P1}, {\bf P2} are obvious. The point {\bf P4} follows
from the definition \eref{e:coeff} of $Q$ and the construction of
$Q_\rho$ in \eref{e:Qrho}.  
To prove {\bf P3}, recall that the map $F=\FR$ of the GL equation is of the type 
\begin{equ}
\FR(u) = \chi\bigl(\|u\|/(3\rho)\bigr) P(u)\;,
\end{equ}
with $P$ some polynomial and $\chi \in \CC_0^\infty(\R)$. The
key point is to notice that the estimate 
\begin{equ}
\|uv\|_\gamma \le C_\gamma\bigl(\|u\|\,\|v\|_\gamma + \|u\|_\gamma\|v\|\bigr)
\end{equ}
holds for every $\gamma \ge 0$, where $uv$ denotes the
multiplication of two functions. In particular, we have 
\begin{equ}
\|u^n\|_\gamma \le C \|u\|_\gamma \|u\|^{n-1}\;,
\end{equ}
which, together with the fact that $\chi$ has compact support, shows
\eref{e:boundF1}. This also shows that the derivatives of $F$ in
$\CH^\gamma$ are polynomially bounded and so \eref{e:boundF2}
holds. {\bf P6} follows by the same argument.

The point {\bf P5} immediately follows from the fact that the image of the operator $\bigl(DQ(x)\bigr)h$ is contained in $\CH_\rL$ for every $x,h \in \CH$.
\end{proof}

\begin{remark}
The condition {\bf P1}
implies that
$e^{-At}$ is an analytic semigroup of contraction operators on $\CHH$.
We will use repeatedly the bound
\begin{equ}
\nn[b]{\gamma }{e^{-At}x}\,\le\,
C_\gamma t^{-\gamma } \nn{}{x}\;.
\end{equ}
\end{remark}

We begin the study of \eref{e:new1} by considering the
equation for the mild solution 
\begin{equa}[e:new2]
\Psi (t, \ic ,\omega )\,&=\,e^{-At} \ic 
+\int_0^t e^{-A(t-s)}F\bigl(\Psi (s, \ic ,\omega )\bigr )\,ds\\
&~~+\int_0^t e^{-A(t-s)}Q\bigl(\Psi (s, \ic ,\omega )\bigr )\,dW(s,\omega )\;.
\end{equa}
The study of this equation is in several steps. We will consider
first the noise term, then the equation for a fixed instance of $\omega $,
and finally prove existence and bounds.

We need some more notation:
\begin{definition}
\label{def:norms}
Let $\CHH^\alpha $ be as above the domain
of $A^\alpha$ with the graph norm.
We fix, once and for all, a maximal time $T$.
We denote by  $\XX \alpha  T $ the space
$\CC([0,T],\CHH^\alpha  )$ equipped with the norm
\begin{equ}
\nn{\XX \alpha T}{y}  = \sup_{t\in[0,T]}  \nn{\alpha }{y(t)} \;.
\end{equ}
We write $\XX{} T$ instead of $\XX 0 T$\,.
\end{definition}
\subsection{The Noise Term}
\label{sec:noise}

Let $y\in \L^p(\Omega,\XX {} T)$. 
(One should think of $y$ as being $y(t)=\Phi^t$.)
The noise term in \eref{e:new2} 
will be studied as a function on $\L^p(\Omega,\XX {} T)$.
It is given by the function $Z$ defined as
\begin{equ}[e:new3]
\bigl(Z(y)\bigr )(\omega ) = t\mapsto \int_0^t
e^{-A(t-s)}Q\bigl(y(\omega)
(s)\bigr )\,dW(s,\omega )\;.
\end{equ}
We will show that $Z(y)$ is in
$\L^p(\Omega ,\XX \alpha T )$ when   $y$ is in $\L^p(\Omega,\XX {} T)$.
The natural norm here is the $\L^p$ norm defined by
\begin{equ}
\nnn{\XX \alpha T ,p}{Z(y)}= \left(\Exp_\omega \sup_{t\in[0,T]}\nn{\alpha
}{(Z(y))_t(\omega )}^p\right)^{1/p}\;.
\end{equ}

\begin{proposition}
\label{prop:noise}
Let $\CHH$, $A$ and $Q$ be as above and assume {\bf P1} and {\bf P4} are
satisfied.
Then, for every $p\ge1$ and every $T < T_0$ one has
\begin{equ}[e:zalone]
\nnn{\XX \alpha  T ,p}{Z(y)}\,\le\,C_{T_0} T^{p/16}\;.
\end{equ}
\end{proposition}
\begin{proof}
Choose an element $y \in \L^p(\Omega,\XX {} T)$. In the sequel, we will consider $y$
as a function over $[0,T]\times\Omega$ and we will not write explicitly the
dependence on $\Omega$.

In order to get bounds on $Z$, we use the factorization formula
and the Young inequality.
Choose $\gamma\in (1/p,1/8)$.
The factorization formula
\cite{ZDP1} then gives the equality
\begin{equ}
\bigl(Z(y)\bigr)(t) = C \int_0^t (t-s)^{\gamma-1} e^{-A(t-s)} \int_0^s
(s-r)^{-\gamma}e^{-A(s-r)}Q(y(r))\,dW(r)\,ds\;.
\end{equ}
Since $A$ commutes with $e^{-At}$, the H\"older inequality leads to
\begin{equs}
{}&\Ann[b]{\alpha }{\bigl(Z(y)\bigr)(t)} ^p \label{e:intermediate}\\
&\,\,= C
\nn[B]{}{
\int_0^t (t-s)^{\gamma-1} e^{-A(t-s)}
\int_0^s (s-r)^{-\gamma}A^\alpha
e^{-A(s-r)}Q(y(r))\,dW(r)\,ds
} ^p
\\
&\,\,\le  C t^{\nu}\int_0^t \Bnn[B]{}{\int_0^s (s-r)^{-\gamma}A^\alpha
e^{-A(s-r)}Q(y(r))\,dW(r)} ^p\,ds\;,
\end{equs}
with $\nu = (p\gamma - 1)/(p-1)$.
For the next bound we need the following result:
\begin{lemma}
\label{lem:estStoch}
\cite[Thm.~7.2]{ZDP1}. Let $r\mapsto\Psi^r$ be an arbitrary predictable $\SL^2(\CHH)$-valued process. Then,
for every $p\ge 2$, there exists a constant $C$ such that
\begin{equ}
\Exp\Bigl( \Bnn[B]{}{\int_0^s \Psi^r\,dW(r)} ^p\Bigr)\le C
\Exp\Bigl(\int_0^s\|\Psi^r\|_\HS^2\,dr\Bigr)^{p/2}\;.
\end{equ}
\end{lemma}
This lemma, the Young inequality applied to
\eref{e:intermediate}, and {\bf P4} above
imply
\begin{equs}
{}&\nnn{\XX \alpha T,p}{Z(y)}^p
= \Exp \Bigl( \sup_{0\le t
\le T} \Bnn[B]{}{\int_0^t\,A^\alpha e^{-A(t-s)}Q\bigl(y(s)\bigr)\,
dW(s)} ^p\Bigr) \\
&~~\le C T^\nu \Exp \int_0^T  \Bnn[B]{}{\int_0^s
(s-r)^{-\gamma} A^{\alpha} e^{-A(s-r)}Q\bigl(y(r)\bigr)\,
dW(r)} ^p\,ds\\
&~~\le C T^\nu \Exp\int_0^T  \Bigl(\int_0^s (s-r)^{-2\gamma}
\nn[b]{\HS}{ A^{\alpha} e^{-A(s-r)}Q\bigl(y(r)\bigr)}^2\,
dr\Bigr)^{p/2}\,ds\\
&~~\le C T^\nu \Exp\int_0^T  \Bigl(\int_0^s (s-r)^{-2\gamma}
\nn[b]{}{ A^{3/8} e^{-A(s-r)}}^2\nn[b]{\HS}{ A^{\alpha-3/8}Q\bigl(y(r)\bigr)}^2\,
dr\Bigr)^{p/2}\,ds\\
&~~\le C T^\nu \Exp\int_0^T\Bigl(\int_0^s (s-r)^{-2\gamma-3/4}
\nn[b]{\HS}{ A^{\alpha-3/8}Q\bigl(y(r)\bigr)}^2\,
dr\Bigr)^{p/2}\,ds\\
&~~\le C T^\nu \Bigl(\int_0^T
s^{-2\gamma-3/4}\,ds\Bigr)^{p/2} \Exp\int_0^T
\bigl\|A^{\alpha-3/8} Q(y(s))\bigr\|_\HS^p\,ds \\
&~~\le C T^{1+\nu} \Bigl(\int_0^T s^{-2\gamma - 3/4}\,ds\Bigr)^{p/2}\;,\label{e:85}
\end{equs}
provided $\gamma < 1/8$. We choose $\gamma = 1/16$ (which thus imposes the condition $p>16$), and we find
\begin{equ}
\nnn{\XX \alpha T,p}{Z(y)}^p \le C T_0^{1+\nu} T^{p/16}\;.
\end{equ}
Thus, we have shown \eref{e:zalone} for $p>16$. Since we are working
in a probability space the case of $p\ge1$ follows. 
This completes the proof of \sref{prop:noise}.
\end{proof}
\subsection{A Deterministic Problem}

The next step in our study of \eref{e:new2} is the analysis of the
problem for a {\em fixed} instance of the noise $\omega $.
Then \eref{e:new2} is of the form 
\begin{equ}
h(t, \ic ,z) = e^{-At}  \ic  + \int_0^t e^{-A(t-s)}F\bigl(h(s, \ic ,z)\bigr
)\,ds+z(t)\;, 
\end{equ}
where we assume that $z\in\CHH^\alpha_T $. One should think of this as
an instance of $Z(\Phi)$, but at this point of our proof, the
necessary bounds are not yet available.

We find it more convenient to study instead of $h$ the quantity $g$
defined by
$g(t, \ic ,z)$ $=h(t, \ic ,z)-z(t)$. Then $g$ satisfies
\begin{equ}[e:new7]
g(t, \ic ,z) = e^{-At}  \ic  + \int_0^t e^{-A(t-s)}F\bigl(g(s, \ic ,z)+z(s)\bigr
)\,ds\;.
\end{equ}
We consider the solution (assuming it exists) as a map from the
initial condition $ \ic $ and the deterministic noise term $z$. More
precisely, we define
\begin{equ}
G( \ic ,z)_t\,=\,g(t, \ic ,z)\;.
\end{equ}
This is a map defined on $\CH\times\CHH^\alpha_T $.
Clearly, \eref{e:new7} reads:
\begin{equ}[e:new77]
G( \ic ,z)_t\,=\,e^{-At} \ic +\int_0^t e^{-A(t-s)}
F\bigl(G( \ic ,z)_s+z(s)\bigr)\,ds\;. 
\end{equ}

To formulate the bounds on $G$, we 
need some more spaces that take into account the regularizing
effect of the semigroup $t\mapsto e^{-At}$. 

\begin{definition}For $\gamma\ge 0 $ 
the spaces $\YY \gamma T  $
are defined as the closures of
$\CC([0,T], \CHH^\gamma )$
under the norm
\begin{equ}
\nn{\YY\gamma T}{y}
= \sup_{t\in(0,T]} t^{\gamma }
\nn{\gamma}{y(t)}
+\sup_{t\in[0,T]}  \nn{}{y(t)}\;.
\end{equ}
\end{definition}
Note that
\begin{equ}
\nn{\YY\gamma T}{y} \,\le\, C_{\gamma ,T}\nn{\XX\gamma T}{y}\;.
\end{equ}
With these definitions, one has:
\begin{proposition}
\label{prop:H14}
Assume the conditions {\bf P1}--{\bf P4} are satisfied.
Assume $ \ic \in\CHH$ and $z\in\XX \alpha T$.
Then, there exists a map $G:\CHH\times\XX \alpha T \to\XX {} T$ 
solving \eref{e:new77}. 
One has the following bounds:
\begin{claim}
\item[\myitem{A}] If $ \ic \in \CHH^\gamma $ with $\gamma \le\alpha $
one has for every $T>0$ the bound
\begin{equ}[e:boundG]
\nn{\XX \gamma T}{ G( \ic ,z)}  \le C(1 +
\nn{\gamma}{ \ic }  + \nn{\XX \gamma T}{z})\;.
\end{equ}
\item[\myitem{B}] If $ \ic \in \CHH $ one has for every $T>0$ the bound
\begin{equ}[e:boundG2]
\nn{\YY\alpha{T}}{ G( \ic ,z)}  \le C(1 +
\nn{}{ \ic }  + \nn{\XX \alpha T}{z})\;.
\end{equ}
\end{claim}
\end{proposition}

Before we start with the proof proper we note the following
regularizing bound:
Define
\begin{equ}[e:better1]
\bigl(\CN f\bigr)(t)\,=\,
\int_0^t e^{-A(t-s)} f(s)\,ds\;.
\end{equ}
Then one has:
\begin{lemma}
\label{lem:better}
For every $\epsilon \in[0,1)$ and every $\gamma>\epsilon $
there is a constant $C_{\epsilon,\gamma} $ such
that
\begin{equ}
\nn{\YY\gamma T}{\CN f}\,\le\,
C_{\epsilon ,\gamma} T \nn{\YY{\gamma-\epsilon}  T}{f}\;,
\end{equ}
for all $f\in\YY{\gamma-\epsilon}  T$.
\end{lemma}
\begin{proof}
We start with
\begin{equs}
\Ann[b]{\gamma}{\bigl(\CN f\bigr)(t)}
\,&\le\,
\int_0^{t/2} \Bnn[b]{}{A^\gamma e^{-A(t-s)} f(s)}\,ds
+
\int_{t/2}^t \Bnn[b]{}{A^\epsilon  e^{-A(t-s)} A^{\gamma -\epsilon
}f(s)}\,ds\\
\,&\le\,
\int_0^{t/2} (t-s)^{-\gamma }\nn{}{ f(s)}\,ds
+
\int_{t/2}^t (t-s)^{-\epsilon }\nn{{\gamma -\epsilon}}{
f(s)}\,ds\\
\,&\le\,
\int_0^{t/2} (t-s)^{-\gamma }\nn{\YY {\gamma -\epsilon} T }{ f}\,ds
+
\int_{t/2}^t (t-s)^{-\epsilon }s^{\epsilon -\gamma }\nn{\YY {\gamma -\epsilon} T }
{f}\,ds\\
\,&\le\,C t^{1-\gamma }\nn{\YY {\gamma -\epsilon} T }
{f}+Ct^{1-\epsilon }t^{\epsilon -\gamma }\nn{\YY {\gamma -\epsilon} T }{f}\;.
\end{equs}
Therefore, $t^\gamma \Ann[b]{\gamma}{\bigl(\CN f\bigr)(t)}\le
C T\nn{\YY {\gamma -\epsilon} T }{f}$. 
Similarly, we have
\begin{equ}
\Bnn[b]{}{\bigl(\CN f\bigr)(t)}\,\le\,\int_0^{t}
\Bnn[b]{}{e^{-A(t-s)} f(s)}\,ds\,\le\,C t\nn{\YY {\gamma -\epsilon} T }{f}\;.
\end{equ}
Combining the two inequalities, the result follows.
\end{proof}
\begin{proof}[of \prop{prop:H14}]
%
%
%
%
We first choose an initial condition $ \ic  \in \CHH^\gamma$ and a function $z
\in \XX \gamma T$.
The local existence of the solutions
in $\CHH^\gamma$ is a well-known result. Thus
there exists, for a possibly small time $\tT>0 $, a function $u\in
\CC([0,\tT ],\CHH^\gamma)$ satisfying
\begin{equ}
u(t) = e^{-At} \ic  + \int_0^t e^{-A(t-s)}F\bigl(u(s) + z(s)\bigr)\,ds\;.
\end{equ}
In order to get an {\it a priori} bound on $\nn{\gamma}{u(t)}$ we
use assumption {\bf P3} and find 
\begin{equs}
\nn{\gamma}{u(t)}& \le\nn{\gamma}{ \ic } + C_{F,\gamma} \int_0^t \bigl(
1
+\nn{\gamma}{u(s)+z(s)}
\bigr)\,ds\\
 &\le C\bigl(1+\nn{\gamma}{ \ic } + \nn{\XX \gamma T}{z}\bigr) +
C_{F,\gamma}  \int_0^t \nn{\gamma}{u(s)}\,ds\;.
\end{equs}
By Gronwall's lemma  we get for $t<T$,
\begin{equ}[e:boundgamma]
\nn{\gamma}{u(t)}  \le  C_1 (1+\nn{\gamma}{ \ic }+\nn{\XX \gamma T}{z})\;.
\end{equ}

Note that \eref{e:boundgamma} tells us that if the initial condition
$ \ic $ is in $\CHH^\gamma$ and if $z$ is in $\XX \gamma T$, then $u(t)$ is, for
small enough $t$, again in  $\CHH^\gamma$ with the above bound.
Therefore, we can iterate the above reasoning and show
the global existence of the solutions up to time $T$, with bounds. 
Thus, $G$ is
well-defined and satisfies the bound
\eref{e:boundG}.

%
%
%
%

We turn to the proof of the estimate \eref{e:boundG2}. 
Define
for $z \in \XX {} T$ the map
$\CM_z$ by
\begin{equ}[e:mdef]
\bigl(\CM_z(x)\bigr)(t) = e^{-At} \ic  + \int_0^t e^{-A(t-s)}F\bigl(x(s) +
z(s)\bigr)\,ds\;.
\end{equ}
Taking $ \ic \in\CHH$ we see from \eref{e:boundgamma} with $\gamma=0$
that
there exists a fixed point $u$ of $\CM_z$ which satisfies
\begin{equ}
\nn{\XX {} T}{u}\,=\,\sup_{t\in[0,T]}\nn{}{u(t)}\,\le\,
 C_1 (1+\nn{}{ \ic }+\nn{\XX {} T}{z})\;.
\end{equ}
Assume next that $z\in \XX \alpha  T$ and hence {\it a fortiori}
$z\in\YY \alpha  T$. 
Then,
by {\bf P3} one has
\begin{equ}
\nn{\YY\gamma T}{F(x+z)}\,\le\,
C \bigl(1 +\nn{\YY\gamma T}{x}+\nn{\YY\gamma T}{z}\bigr)\;.
\end{equ}
Since $u$ is a fixed point and \eref{e:mdef} contains a term of the
form of \eref{e:better1} we can apply \lem{lem:better} and obtain for
every $\gamma \le \alpha $ and $\epsilon \in[0,1)$:
\begin{equs}
\nn{\YY {\gamma +\epsilon }  T}{u}
\,&=\,\nn{\YY {\gamma +\epsilon }  T}{\CM_z(u)}
\,\le\,C\nn{}{ \ic }+ CT \nn{\YY\gamma T}{F(u+z)}\\
\,&\le\,C\nn{}{ \ic }+C \bigl(1 +\nn{\YY\gamma T}{u}+\nn{\YY\gamma
T}{z}\bigr)\;.\label{e:nochbesser} 
\end{equs}
Thus, as long as $\nn{\YY\gamma T}{z}$ is finite, we can apply
repeatedly
\eref{e:nochbesser} until reaching $\gamma =\alpha $, and this proves
\eref{e:boundG2}.  
The proof of \sref{prop:H14} is complete.
\end{proof}

\subsection{Stochastic Differential Equations in Hilbert Spaces}
Before we can start with the final steps of the proof of
\sref{prop:reg} we state in the next subsection a general existence theorem for
stochastic differential equations in Hilbert spaces.
The symbol $\CHH$ denotes
a separable Hilbert space. We are interested in solutions to the SDE
\begin{equ}[e:genSDE]
dX^t = (-AX^t  + N(t,\omega ,X^t ) + M^t)\,dt + B(t,\omega ,X^t )\, dW(t)\;,
\end{equ}
where $W(t)$ is the cylindrical Wiener process on a separable Hilbert space
$\CHH_0$. We assume $B(t,\omega ,X^t ):\CH_0\to \CH$ is
Hilbert-Schmidt.
We will denote by $\Omega$ the underlying probability space and by
$\{\cscr F_t\}_{t\ge 0}$ the associated filtration.

The exact conditions spell out as follows:
\begin{claim}
\item[{\bf C1}] The operator $A : \CD(A) \to \CHH$ is the generator of a strongly
continuous semigroup in $\CHH$.
\item[{\bf C2}] There exists a constant $C>0$ such that for arbitrary $x,y \in
\CHH$, $t \ge 0$ and $\omega \in \Omega$ the estimates
\begin{equs}[0]
\nn{}{N(t,\omega,x) - N(t,\omega, y)} + 
\|B(t,\omega,x) - B(t,\omega,y)\|_\HS \le
C
\nn{}{x-y}\;, \\
\nn{}{N(t,\omega,x)}^2 + 
\|B(t,\omega,x)\|_\HS^2 \le C^2(1+
\nn{}{x}^2)\;,
\end{equs}
hold.
\item[{\bf C3}] For arbitrary $x,h \in \CHH$ and $h_0 \in \CHH_0$, the stochastic
processes $\scal{N(\cdot,\cdot,x),h}$ and $\scal{B(\cdot,\cdot,x)h_0,h}$ are
predictable.
\item[{\bf C4}] The $\CHH$-valued stochastic process $M^t$ is predictable, has continuous sample paths, and satisfies
\begin{equ}
\sup_{t\in[0,T]} \Exp\; \nn{}{M^t}^p < \infty\;,
\end{equ}
for every $T>0$ and every $p\ge 1$.
\item[{\bf C5}] For arbitrary $t>0$ and $\omega \in \Omega$, the maps $x \mapsto
N(t,\omega,x)$ and $x \mapsto B(t,\omega,x)$ are twice continuously
differentiable with their derivatives bounded by a constant independent of $t$,
$x$ and $\omega$.
\end{claim}
We have the following existence theorem.
%
%
\begin{theorem}
\label{theo:exist}
Assume that $ \ic \in\CHH$
and that {\bf C1} -- {\bf C4}~are satisfied.
\begin{claim}
\item[--] For any $T>0$, there exists a mild solution
$X_ \ic ^t$
of \eref{e:genSDE} with $X_ \ic ^0 =  \ic $. 
This solution is unique among the $\CHH$-valued processes satisfying
\begin{equ}
\Prob\left (  \int_0^T \Bnn[b]{}{X^t_ \ic }^2\,dt<\infty  \right )\,=\,1\;.
\end{equ}
Furthermore, $X_ \ic $ has a
continuous version and is strongly Markov.
\item[--] For every $p \ge 1$ and $T>0$, there exists a constant $ C_{p,T}$
such that
\begin{equ}[e:exist99]
\Exp \bigl(\sup_{t \in[0,T]}\Bnn[b]{}{X^t_ \ic }^p\bigr) \le  C_{p,T} (1 +
\nn{}{ \ic }^p)\;.
\end{equ}
\item[--] If, in addition, {\bf C5}~is satisfied, the mapping $ \ic 
\mapsto X_ \ic ^t(\omega)$ 
has a.s.~bounded partial derivatives with respect to the initial
condition $ \ic $. These derivatives satisfy the SDE's obtained by formally
differentiating \eref{e:genSDE} with respect to $X$.
\end{claim}
\end{theorem}
\begin{proof}
The proof of this theorem for the case $M^t \equiv 0$ can be found in
\cite{ZDP}. The same proof carries through for the case of non-vanishing
$M^t$ satisfying {\bf C4}. 
\end{proof}
%
\subsection{Bounds on the Cutoff Dynamics (Proof of \prop{prop:reg})}
\label{sec:propreg}
%
With the tools from stochastic analysis in place, we can now prove 
\sref{prop:reg}. 
We start with the

%
%
\begin{likerem}[Proof of (A).]
In this case we identify the equation \eref{e:genSDE} with
\eref{e:SGLrho} and apply \theo{theo:exist}.
The condition {\bf C1} of \theo{theo:exist} is
obviously true, and the 
condition 
{\bf C3} is redundant in this case.
The condition {\bf C2} is satisfied because $F$ and $Q$ of
\eref{e:genSDE} 
satisfy {\bf P2}--{\bf P4}. Therefore, \eref{e:exist99} holds and hence
we have shown \eref{e:boundPhi2} for the case of $\gamma =0$.
In particular, $\PhiR^t$ exists and satisfies
\begin{equa}[e:noch1]
\PhiR^t( \ic ,\omega )\,&=\, e^{-At} \ic  +\int_0^t e^{-A(t-s)} F\bigl(
\PhiR^s( \ic ,\omega )\bigr)\,ds\\&~~+\int_0^t e^{-A(t-s)} Q\bigl(
\PhiR^s( \ic ,\omega )\bigr)\,dW(s)\;.
\end{equa}

We can extend \eref{e:boundPhi2} to arbitrary $\gamma \le\alpha $ as follows.
We set as in \eref{e:new3},
\begin{equ}[e:noch11]
\bigl(Z(\PhiR )\bigr)_t(\omega )\,=\,\int_0^t e^{-A(t-s)} Q\bigl(
\PhiR^s( \ic ,\omega )\bigr)\,dW(s)\;.
\end{equ}
By \sref{prop:noise}, we find that for all $p\ge1$ one has
\begin{equ}[e:noch2]
\left(\Exp_\omega \sup_{t\in[0,T]}\nn{\alpha
}{(Z(\PhiR))_t(\omega )}^p\right)^{1/p}<C_{T,p}
\end{equ}
for all $ \ic $. From this, we conclude that, almost surely,
\begin{equ}[e:noch3]
\sup_{t\in[0,T]}\nn{\alpha
}{(Z(\PhiR))_t(\omega )}<\infty \;.
\end{equ}
%
Subtracting \eref{e:noch11} from \eref{e:noch1} we get
\begin{equa}[0][e:noch234]
\PhiR^t( \ic ,\omega )-\bigl(Z(\PhiR )\bigr)_t(\omega )\,=\, e^{-At} \ic  +\int_0^t e^{-A(t-s)} F\bigl(
\PhiR^s( \ic ,\omega )\bigr)\,ds
\\
\,=\, e^{-At} \ic  +\int_0^t e^{-A(t-s)} F\Bigl(
\PhiR^s( \ic ,\omega )-\bigl(Z(\PhiR )\bigr)_s(\omega
)+\bigl(Z(\PhiR )\bigr)_s(\omega )\Bigr)\,ds\;.
\end{equa}
Comparing \eref{e:noch234} with \eref{e:new77} we see that, {a.s.},
\begin{equ}
\PhiR^t( \ic ,\omega )-\bigl(Z(\PhiR )\bigr)_t(\omega )=G\bigl( \ic ,Z\bigl(\PhiR( \ic ,\cdot)\bigr)(\omega )\bigr)\;.
\end{equ}
%
%
We now use $z$ as a shorthand:
\begin{equ}
z(t)\,=\,\Bigl(Z\bigl(\PhiR( \ic ,\cdot)\bigr)\Bigr)_t(\omega )\;.
\end{equ}
Assume now $ \ic \in\CH^\gamma $. Note that by \eref{e:noch3}, $z(t)$ is
in $\CH^\alpha $. If $\gamma \le\alpha $, 
we can apply
\sref{prop:H14} and from \eref{e:boundG} we conclude that
almost surely,
\begin{equ}
\sup_{t\in[0,T]}\nn{\gamma}{ G( \ic ,z)}  \le C\bigl(1 +
\nn{\gamma}{ \ic }  + \sup_{t\in[0,T]}\nn{\gamma}{z}\bigr)\;.
\end{equ}

Finally, since $\gamma \le\alpha $, we find
\begin{equs}
\Exp\biggl(
\sup_{t\in[0,T]}\nn{\gamma 
}{\PhiR^t( \ic  )}^p
\biggr)\,&\le\,C
\Exp\biggl(\sup_{t\in[0,T]}\nn{\gamma 
}{G( \ic ,z)_t}^p
\biggr)+
C\Exp\biggl(\sup_{t\in[0,T]}\nn{\gamma 
}{z(t)}^p
\biggr)\\
\,&\le\,C(1+\nn{\gamma }{ \ic })^p+C\Exp\biggl(\sup_{t\in[0,T]}\nn{\gamma 
}{z(t)}^p
\biggr)\\
\,&\le\,C(1+\nn{\gamma }{ \ic })^p\;, \label{e:inequal}
\end{equs}
where we applied \eref{e:noch2}  to get the last inequality.
Thus, we have shown \eref{e:boundPhi2} for all $\gamma \le\alpha $.
The fact that the solution is strong if $\gamma \ge 1$ is an immediate
consequence of \cite[Lemma 4.1.6]{Lun} and \cite[Thm.~5.29]{ZDP1}.
\end{likerem}
%
%
\begin{likerem}[Proof of (B).]
This bound can be shown in a similar way,
using the bound \eref{e:boundG2} of \sref{prop:H14}: Take $ \ic  \in
\CH$. By the 
above, we 
know that there exists a solution to \eref{e:noch1} satisfying the
bound \eref{e:boundPhi1} with $\gamma = 0$. We define $z(t)$ and
$G( \ic ,z)_t$ as above. But now we apply the bound \eref{e:boundG2}
of
\sref{prop:H14} and we conclude that
almost surely,
\begin{equ}
\sup_{t\in[0,T]} t^\alpha \nn{\alpha}{ G( \ic ,z)}  \le C\bigl(1 +
\nn{}{ \ic }  + \sup_{t\in[0,T]}\nn{\alpha}{z}\bigr)\;.
\end{equ}
Following a procedure similar to \eref{e:inequal}, we conclude that \eref{e:boundPhi1} holds. 
\end{likerem}
%
%
\begin{likerem}[Proof of (C).]The existence of the
partial derivatives follows from \theo{theo:exist}. To show the bound,
choose $ \ic \in \CH$ and $h \in \CH$ with $\|h\| = 1$, and define the
process $\Psi^t = \bigl(D\PhiR^t( \ic )\bigr)h$. It is by
\theo{theo:exist} a mild solution to the equation 
\begin{equ}[e:jacPhi] 
d\Psi^t = -A \Psi^t\, dt + \Bigl(\bigl(DF \circ \PhiR^t\bigr)( \ic )
\Psi^t\Bigr)\,dt + \Bigl(\bigl(DQ \circ \PhiR^t\bigr)( \ic )
\Psi^t\Bigr)\,dW(t)\;. 
\end{equ}
By {\bf P3} and {\bf P5}, this equation satisfies conditions {\bf
C1}--{\bf C3} of \theo{theo:exist}, so we can apply it to get the
desired bound \eref{e:boundDPhi1}. (The constant term drops since the
problem is linear in $h$.)
\end{likerem}
%
%
\begin{likerem}[Proof of (D).]Choose $h\in \CH$ and $ \ic  \in \CH^\alpha$
and define as above $\Psi^t = \bigl(D\PhiR^t( \ic )\bigr)h$, which is
the mild solution to \eref{e:jacPhi} with initial condition $h$. We
write this as 
\begin{equa}
\Psi^t &= e^{-At}h + \int_0^t e^{-A(t-s)} \Bigl(\bigl(DF \circ
\PhiR^s\bigr)( \ic ) \Psi^s\Bigr)\,ds \\ 
&\quad + \int_0^t e^{-A(t-s)} \Bigl(\bigl(DQ \circ
\PhiR^s\bigr)( \ic ) \Psi^s\Bigr)\,dW(s) \\ 
&\equiv S_1^t + S_2^t + S_3^t\;.
\end{equa}
The term $S_1^t$ satisfies
\begin{equ}[e:t1]
\sup_{t\in(0,T]}t^\alpha \nn{\alpha }{S_1^t}\,\le\,C\nn{}{h}\;.
\end{equ}
The term $S_3^t$ is very similar to what is found in \eref{e:new3},
with $Q\bigl( y(s)\bigr)$ replaced by $(DQ\circ \Phi^s_\rho) \Psi^s$.
Repeating the steps of \eref{e:85} for a sufficiently large $p$,
we obtain now with $\gamma={1\over 4}$, some $\mu>0$ and
writing $X^s=\bigl(DQ \circ
\PhiR^s\bigr)( \ic ) \Psi^s$:
\begin{equs}
{}\Exp\sup_{t\in[0,T]}& \nn{\alpha }{S_3^t}^p
= \Exp \Bigl( \sup_{0\le t
\le T} \Bnn[B]{}{\int_0^t\,A^\alpha e^{-A(t-s)}X^s\,
dW(s)} ^p\Bigr) \\
&~~\le C T^\mu \Exp \int_0^T  \Bnn[B]{}{\int_0^s
(s-r)^{-\gamma} A^{\alpha} e^{-A(s-r)}X^r\,
dW(r)} ^p\,ds\\
&~~\le C T^\mu \Exp\int_0^T  \Bigl(\int_0^s (s-r)^{-2\gamma}
\nn[b]{\HS}{ A^{\alpha} e^{-A(s-r)}X^r}^2\,
dr\Bigr)^{p/2}\,ds\\
&~~\le C T^\mu \Exp\int_0^T  \Bigl(\int_0^s (s-r)^{-2\gamma}
\nn[b]{\HS}{ A^{\alpha}X^r}^2\,
dr\Bigr)^{p/2}\,ds\\
&~~\le C T^\mu \Bigl(\int_0^T
s^{-2\gamma}\,ds\Bigr)^{p/2} \Exp\int_0^T
\bigl\|A^{\alpha}X^s \bigr\|_\HS^p\,ds \\
&~~\le C T^{\mu+p/4} \Exp\int_0^T
\bigl\|A^{\alpha}\bigl(DQ \circ
\PhiR^s\bigr)( \ic ) \Psi^s \bigr\|_\HS^p\,ds\;.
\end{equs}
We now use {\bf P5}, \ie, \eref{e:P5} and then \eref{e:boundDPhi1}  
and get
\begin{equ}[e:t3]
\Exp\sup_{t\in[0,T]} \nn{\alpha }{S_3^t}^p
\,\le\,C T^{\mu+p/4}\Exp \int_0^T
\nn{}{\Psi^s}^p\,ds\,\le\,C T^{\mu+p/4+1} \nn{}{h}^p\;.
\end{equ}

To treat the term $S_2^t$, we fix a realization
$\omega\in \Omega$ of the noise and use \lem{lem:better}. This gives
for $\eps \in [0,1)$ the bound 
\begin{equ}
\sup_{t\in (0,T]} t^\gamma \nn{\gamma}{S_2^t} \le C T \sup_{t\in (0,T]} t^{\gamma - \eps} \nn[b]{\gamma-\eps}{\bigl(DF \circ \PhiR^t\bigr)( \ic ) \Psi^t}\;.
\end{equ}
By {\bf P6}, this leads to the bound, a.s.,
\begin{equ}
\sup_{t\in (0,T]} t^\gamma \nn{\gamma}{S_2^t} \le C_T\Bigl(1+\sup_{t\in (0,T]} \nn[b]{\gamma-\eps}{\PhiR^t( \ic )}\Bigr) \sup_{t\in (0,T]} t^{\gamma - \eps} \nn[b]{\gamma-\eps}{\Psi^t}\;.
\end{equ}
Taking expectations we have
\begin{equ}
\Exp\sup_{t\in (0,T]} t^{\gamma p } \nn{\gamma}{S_2^t}^p \le
C_T^p\Exp\left (
\Bigl(1+\sup_{t\in (0,T]} \nn[b]{\gamma-\eps}{\PhiR^t( \ic )}\Bigr)^p
\sup_{t\in (0,T]} t^{(\gamma - \eps)p}
\nn[b]{\gamma-\eps}{\Psi^t}^p\right )\;.
\end{equ}
By the Schwarz inequality and \eref{e:boundPhi2} we get
\begin{equ}[e:t2]
\Exp\sup_{t\in (0,T]} t^{\gamma p } \nn{\gamma}{S_2^t}^p \le
C 
\bigl(1+ \nn{\gamma-\eps}{ \ic }^p\bigr)
\Bigl(\Exp\sup_{t\in (0,T]} t^{(\gamma - \eps)2p}
\nn[b]{\gamma-\eps}{\Psi^t}^{2p}\Bigr)^{1/2}\;.
\end{equ}
Since $\Psi^t=\bigl(D\PhiR ^t( \ic )\bigr)h=S_1^t+S_2^t+S_3^t$,
combining \eref{e:t1}--\eref{e:t2} leads to
\begin{equs}
\Exp\sup_{t\in (0,T]}& t^{\gamma p } \nn{\gamma}{\bigl(D\PhiR
^t( \ic )\bigr)h}^p\\& \le C\nn{}{h}^p+
C 
\bigl(1+ \nn{\gamma-\eps}{ \ic }^p\bigr)
\Bigl(\Exp\sup_{t\in (0,T]} t^{(\gamma - \eps)2p}
\nn[b]{\gamma-\eps}{\bigl(D\PhiR
^t( \ic )\bigr)h}^{2p}\Bigr)^{1/2}\;.
\end{equs}
Thus, we have gained $\epsilon $ in regularity. 
Choosing $\epsilon =\HALF$ and
iterating sufficiently
many times we obtain \eref{e:boundDPhi2} for sufficiently large $p$. The
general case then follows from the H\"older inequality.
\end{likerem}
%
%
\begin{likerem}[Proof of (E).]We estimate this expression by
\begin{equ}
\nn[b]{\gamma}{\PhiR^t( \ic ) - e^{-At} \ic } \le \int_0^t \nn[b]{\gamma}{F\bigl(
\PhiR^s( \ic )\bigr)}\,ds + \nn[bb]{\gamma}{\int_0^t e^{-A(t-s)} Q\bigl(
\PhiR^s( \ic )\bigr)\,dW(s)}\;.
\end{equ}
The first term can be bounded by combining \sref{e:boundPhi1} and {\bf P3}. The second term is bounded by \sref{prop:noise}.
\end{likerem}
The proof \sref{prop:reg} is complete.

\subsection{Bounds on the Off-Diagonal Terms}
\label{sec:cross}

Here, we prove \lem{lem:cross}. This is very similar to the proof of
\myitem{D} of \sref{prop:reg}. 
\begin{proof}
We start with \eref{e:cross2}. Recall that here we do not write the
cutoff $\rho $.
We choose $h\in\CH_\rH$ and $ \ic \in\CH$. The equation for $\Psi^t=\bigl(D_\rH
\Phi^t_{\rL}( \ic )\bigr)h$ is : 
\begin{equa}
\Psi^t &= 
\int_0^t e^{-A(t-s)} \Bigl(\bigl(D F_\rL \circ
\PhiR^s\bigr)( \ic ) \bigl(D_\rH
\Phi^s( \ic )\bigr)h\Bigr)\,ds \\ 
&~~+ \int_0^t e^{-A(t-s)}
\Bigl(\bigl(D Q_\rL \circ
\PhiR^s\bigr)( \ic ) \bigl(D_\rH
\Phi^s( \ic )\bigr)h\Bigr)\,dW(s) \\ 
&\equiv  R_1^t + R_2^t\;.
\end{equa}
Since $DF=D\FR$ is bounded
we get
\begin{equa}
\nn[b]{}{R_1^t}
\,\le\,
C\int_0^t \nn[b]{}{\bigl(D_\rH
\Phi^s( \ic )\bigr)h}\,ds
\,\le\,
Ct\sup_{s\in[0,t]} \nn[b]{}{\bigl(D_\rH
\Phi^s( \ic )\bigr)h}\;.
\end{equa}
Using \eref{e:boundDPhi1}, this leads to
\begin{equ}
\Exp\sup_{t\in[0,T]}\nn[b]{}{R_1^t}
^p\,\le\,
C^pt^p\Exp\sup_{s\in[0,t]} \nn[b]{}{\bigl(D_\rH
\Phi^s( \ic )\bigr)h}^p\,\le\,Ct^p\nn{}{h}^p\;.
\end{equ}
The term $R_2^t$ is bounded exactly as in \eref{e:t3}.
Combining the bounds, \eref{e:cross2} follows.

Since $Q_\rH$ is constant, see \eref{e:Qrho}, we get for  
$\Psi^t=\bigl(D_\rL
\Phi^t_{\rH}( \ic )\bigr)h$ and $h\in\CH_\rL$: 
\begin{equa}
\Psi^t &= 
\int_0^t e^{-A(t-s)} \Bigl(\bigl(D F_\rH \circ
\PhiR^s\bigr)( \ic ) \bigl(D_\rL
\Phi^s( \ic )\bigr)h\Bigr)\,ds\;.
\end{equa}
This is bounded like $R_1^t$ and leads to \eref{e:cross1}.
This completes the proof of \lem{lem:cross}.
\end{proof}
\subsection{Proof of \prop{prop:allerlei}}
\label{sec:theend}

Here we point out where to find the general results on \eref{e:SGLL}
which we stated in \sref{prop:allerlei}.
Note that these are bounds on the flow {\em without} cutoff
$\rho $.
\begin{proof}[of \prop{prop:allerlei}]There are many ways to prove
this. To make things simple, 
without getting the best estimate possible, we note that a bound in
$\L^\infty $ can be found in \cite[Prop.~3.2]{Cerr}. To get from
$\L^\infty $ to $\CH$, we note that $ \ic \in\CH$ and we use \eref{e:SGLL}
in its integral form.
The term
$e^{-At} \ic $ is bounded in $\CH$, while the non-linear term $\int_0^t
e^{-A(t-s)}F\bigl(\Phi^s( \ic )\bigr)\,ds$ can be bounded by using a
version of \lem{lem:better}. Finally, the noise term is bounded by \sref{prop:noise}.

Furthermore, because of the
compactness of the semigroup generated by $A$, it is possible to show
\cite[Thm.~6.3.5]{ZDP} that an invariant measure exists. 
\end{proof}

\subsection*{Acknowledgements}We thank L. Rey-Bellet and G. Ben-Arous 
for helpful discussions. This research was partially supported by the
Fonds National Suisse.
\bibliographystyle{myalph}
\markboth{\sc \refname}{\sc \refname}
\def\Rom#1{\uppercase\expandafter{\romannumeral #1}}
\providecommand{\bysame}{\leavevmode\hbox to3em{\hrulefill}\thinspace}

\end{document}